\documentclass[fleqn,usenatbib]{mnras}

\usepackage{newtxtext,newtxmath}
\usepackage[T1]{fontenc}

\DeclareRobustCommand{\VAN}[3]{#2}
\let\VANthebibliography\thebibliography
\def\thebibliography{\DeclareRobustCommand{\VAN}[3]{##3}\VANthebibliography}

\usepackage{graphicx}	% Including figure files
\usepackage{stfloats}
\usepackage{float}
\usepackage{lipsum}
\usepackage[normalem]{ulem}
\usepackage{subfig}
\usepackage[justification=centering]{caption}
\usepackage{hyperref}
\usepackage{xfrac}

\newcommand{\RNum}[1]{\uppercase\expandafter{\romannumeral #1\relax}}
\newcommand{\msun}{M\textsubscript{\(\odot\)}}
\newcommand{\lsun}{L\textsubscript{\(\odot\)}}
\newcommand{\mjup}{M\textsubscript{J}}

\title[Kinematic excess, gas rings and cavity]{A kinematic excess in the annular gap and gas depleted cavity in the disc around HD\,169142}

\author[H. Garg et al.]{
H. Garg$^{1}$\thanks{E-mail: himanshi.garg@monash.edu}, C. Pinte$^{1,2}$, I. Hammond$^{1}$, R. Teague$^{3,4}$, T. Hilder$^{1}$, D.~J. Price$^{1}$, J. Calcino$^{5}$, V. Christiaens$^{6}$, \newauthor P.~P. Poblete$^{7,8}$
\\
$^{1}$School of Physics and Astronomy, Monash University, Clayton VIC 3800, Australia\\
$^{2}$Univ. Grenoble Alpes, CNRS, IPAG, F-38000 Grenoble, France\\
$^{3}$Department of Earth, Atmospheric, and Planetary Sciences, Massachusetts Institute of Technology, Cambridge, MA 02139, USA\\
$^{4}$Center for Astrophysics $|$ Harvard \& Smithsonian, 60 Garden Street, Cambridge, MA 02138, USA\\
$^{5}$Theoretical Division, Los Alamos National Laboratory, Los Alamos, NM 87545, USA\\
$^{6}$Space sciences, Technologies \& Astrophysics Research (STAR) Institute, Universit\'e de Li\`ege, All\'ee du Six Ao\^ut 19c, B-4000 Sart Tilman, Belgium\\
$^{7}$Astrophysikalisches  Institut, Friedrich-Schiller-Universit\"at Jena, Schillerg\"aßchen 2–3, 07745 Jena, Germany\\
$^{8}$N\'ucleo Milenio de Formaci\'on Planetaria (NPF), Chile\\
}

\date{Accepted XXX. Received YYY; in original form ZZZ}

\pubyear{2022}

\begin{document}
\label{firstpage}
\pagerange{\pageref{firstpage}--\pageref{lastpage}}
\maketitle

% Abstract of the paper
\begin{abstract}
We present ALMA band 6 images of the $^{12}$CO, $^{13}$CO and C$^{18}$O $J$=2-1 line emissions for the circumstellar disc around HD\,169142, at $\sim$8\,au spatial resolution. We resolve a central gas depleted cavity, along with two independent near-symmetric ring-like structures in line emission: a well-defined inner gas ring [$\sim$25\,au] and a second relatively fainter and diffuse outer gas ring [$\sim$65\,au]. We identify a localised super-Keplerian feature or vertical flow with a magnitude of $\sim$75ms$^{-1}$ in the $^{12}$CO map. This feature has the shape of an arc that spans azimuthally across a PA range of -60$^{\circ}$ to 45$^{\circ}$ and radially in between the B1[26au] and B2[59au] dust rings. Through reconstruction of the gas surface density profile, we find that the magnitude of the background perturbations by the pressure support and self-gravity terms are not significant enough to account for the kinematic excess. If of planetary origin, the relative depletion in the gas-density profile would suggest a 1\,M$_{\mathrm{J}}$ planet. In contrast, the central cavity displays relatively smooth kinematics, suggesting either a low mass companion and/or a binary orbit with a minimal vertical velocity component.
\end{abstract}

% Select between one and six entries from the list of approved keywords.
% Don't make up new ones.
\begin{keywords}
accretion discs -- circumstellar matter -- submillimetre: planetary systems -- stars: individual: HD169142
\end{keywords}

%%%%%%%%%%%%%%%%%%%%%%%%%%%%%%%%%%%%%%%%%%%%%%%%%%

%%%%%%%%%%%%%%%%% BODY OF PAPER %%%%%%%%%%%%%%%%%%

\section{Introduction}

High resolution imaging of discs around pre-main sequence stars by the ALMA (Atacama Large Millimeter/sub-millimetre Array) interferometer have revealed sub-structures in the form of gaps and rings primarily seen in dust emission \citep[e.g.][]{ALMApartnership, Andrews2016, Andrews2018, Sierra2021}. One popular mechanism for the formation of concentric rings is the dynamical interaction of the disc with embedded planets \citep[e.g.][]{Ayliffe2012, Pinilla2012, Dipierro2015, Rosotti2016, Dong2016, Dong2017, Veronesi2020}. Alternatives include, magneto-hydrodynamic instabilities \citep[e.g.][]{Flock2015, Pinilla2016} and condensation fronts \citep[e.g.][]{Kretke2007, Saito2011, Zhang2015, Okuzumi2016}.

Embedded planets have long been predicted to perturb the gas component of the disc \citep{Rafikov2002}. The gravitational force exerted by the planet on the natal disc results in the formation of density waves launched at Lindblad resonances that superimpose to resemble a coherent wake propagating away from the location of the planet. This translates to local deviations in the otherwise uniform velocity fields, that gradually diminish in magnitude with increasing distance from the planet. The magnitude of the deviations are directly correlated to the mass of the perturbing planet \citep[e.g.][]{Bollati2021}.

Empirical methods based on gas kinematics are becoming common practice to indirectly infer the presence of an embedded planet \citep{Perez2018}. Planets can carve gaps in the gas density profile and the associated pressure gradients can manifest as velocity perturbations which can indirectly reveal the presence of the planets \citep{Teague2018}. Specifically, the local density depletion equates to negative and positive pressure gradients along the boundaries, which translate to sub- and super-Keplerian rotation with respect to the background Keplerian profile, respectively. Whilst this observation is independent of the underlying mechanism responsible for the generation of the gaps, velocity deviations closely resembling a wake, favour a planetary origin. Deviations in velocity channels ("kinks") from the expected isovelocity curves can also be used to infer the presence of a planet, especially when the location of the gas signature correlates well with the dust gaps \citep[e.g.][]{Pinte2019}. Acquiring the gas surface density profile allows for the computation of the background rotational velocity profile for a pressure supported disc. Furthermore, as an alternative approach to kinematics, the magnitude of the relative depletion within a dust gap can also be directly correlated to planetary masses expected to carve such a gap \citep[e.g.][]{Kanagawa2015, Dong2017, Bae2018, Lodato2019, Bollati2021}.

HD\,169142 is a Herbig Ae star with an estimated mass of M$_{*}$=1.65\msun \citep{Blondel2006}, luminosity L$_{*}=10$\lsun \citep{Fedele2017} and an effective temperature of T$_{*}$=8400K \citep{Dunkin1997}, located at a distance of 117\,$\pm$\,4\,pc away \citep{Gaia2016}. The disc encircling this star is oriented nearly face-on at an inclination of 13$^{\circ}$ \citep{Raman2006, Panic2008} and a position angle (PA) of 5$^{\circ}$ for the major axis \citep{Raman2006}. This disc is one of many to comprise of distinct ring-like structures (r$\sim$25au and $\sim$65\,au) and a central cavity (R$_{\mathrm{cav}}\sim$22au) as seen in scattered light \citep{Quanz2013, Momose2015, Pohl2017, Bertrang2018}; thermal mid-infrared \citep{Honda2012}; millimetre/sub-millimetre with ALMA \citep{Fedele2017, Macias2019, Perez2019}; and centimetre with the VLA (Very Large Array) \citep{Osorio2014}. At higher spatial resolution the outer ring was resolved into three independent rings with radial separations of $\sim$10au \citep{Perez2019}. High-contrast imaging has revealed various point-like sources along the inner and outer edges of the inner-most ring \citep{Biller2014, Reggiani2014, Ligi2018, Gratton2019}, but the close overlap of these features with the dust ring prevented an unequivocal conclusion on the presence of embedded planets \citep{Biller2014, Ligi2018}. Exterior to the dust rings, signatures of meridional flows have also been reported at a radius of 125\,au \citep{Yu2021}.

In this paper, we present ALMA band 6 observations of the disc around HD\,169142 for the $^{12}$CO, $^{13}$CO and C$^{18}$O $J$=2-1 line transitions imaged at 0\farcs07 and 0\farcs1 angular resolutions and 0.167 km\,s$^{-1}$ spectral resolution, taking into account Hanning Smoothing by the correlator. This level of angular resolution allowed us to resolve concentric gas rings in all three line tracers and detect small scale ($\sim$50\,m/s) velocity perturbations stemming from in between the dust rings.

%--------------------------------------------------
\begin{figure*}
    \centering
    \includegraphics[width=\textwidth]{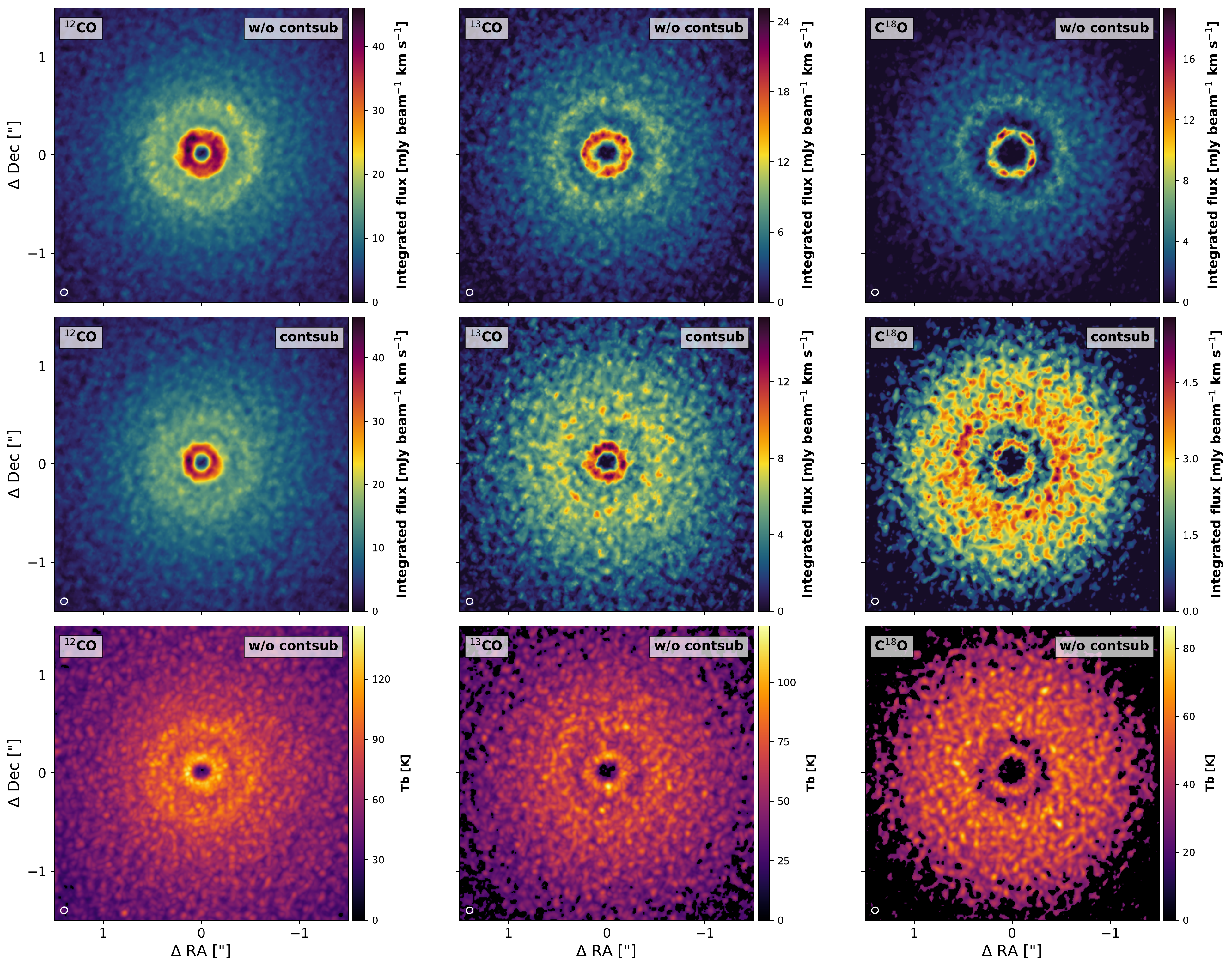}
    \caption{\textbf{Gas rings, gaps and cavity}. Integrated intensity and peak intensity maps (given in units of brightness temperature) for the $^{12}$CO, $^{13}$CO and C$^{18}$O $J$=2-1 maps, with and without continuum subtraction. The disc is inclined by 13$^{\circ}$ perpendicular to a PA of 5$^{\circ}$. All maps are imaged at a resolution of 0\farcs07 with Briggs robustness factor of 0.0 and a uv-taper to circularise the beam, which is displayed on the bottom-left corner.}
    \label{fig:COmaps,intensity}
\end{figure*}

\begin{figure*}
    \centering
    \includegraphics[width=\textwidth]{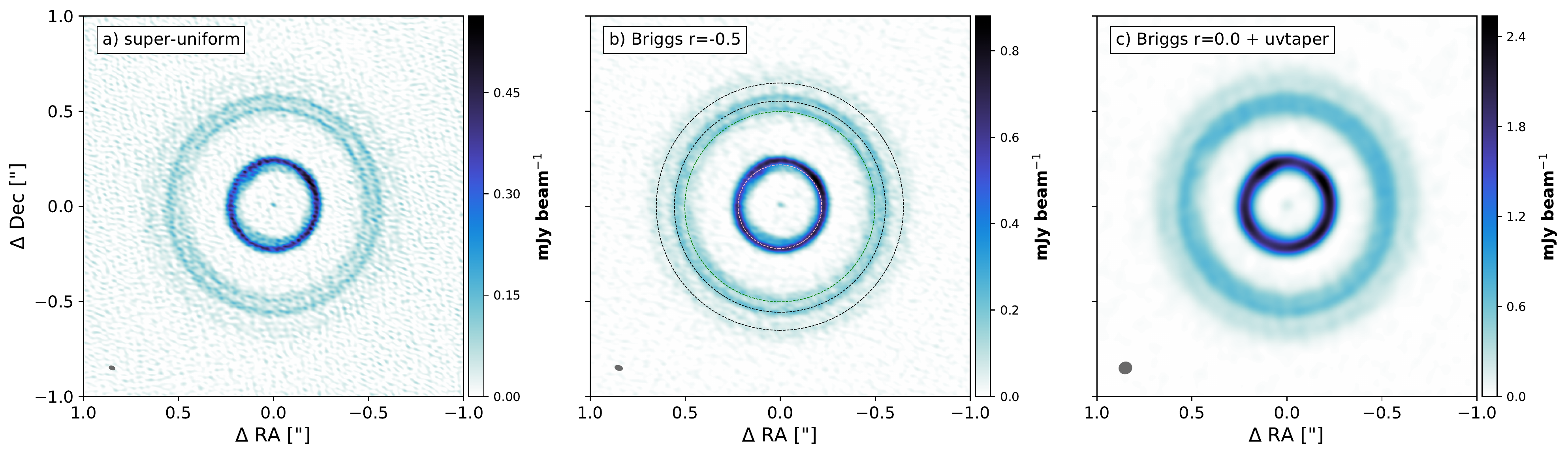}
    \caption{\textbf{Thermal continuum emission.} a) super-uniform weighted map; b) slightly degraded resolution with Briggs r=-0.5. The overlaid dotted rings represent the locations of the B1[26au], B2[59au], B3[66au] and B4[77au] dust rings as presented in \citet{Perez2019} ; c) continuum emission imaged at a similar spatial resolution to line emission ($\sim$0\farcs07).}
    \label{fig:dustmaps}
\end{figure*}

\section{Observations, data reduction and imaging}

\begin{table*}
    \caption{Summary of the ALMA observations used for imaging HD\,169142.}
    \resizebox{\textwidth}{!}{%
    \begin{tabular}{lcllcclcccc}
    \hline
    \hline
    Project & P.I. & Date & Execution & Source Int. & \textit{N}$_\mathrm{ant}$ & Baselines & Continuum & Flux & Bandpass & Phase \\
    code & & & & time (mins) & & (m) & CtrFreq, TotBW (GHz) & cal. & cal. & cal. \\
    \hline
    [2015.1.00490.S] & M. Honda & 2016 Sep 14 & 1 & 49.6 & 38 & 15 - 3200 & 232.966, 1.875 & J1733-1304 & J1924-2914 & J1820-2528 \\
    & & 2016 Sep 14 & 2 & 49.6 & 38 & 12.5 - 2950 & 232.966, 1.875 & J1733-1304 & J1924-2914 & J1820-2528 \\
    & & 2016 Sep 14 & 3 & 49.6 & 38 & 14.5 - 2650 & 232.966, 1.875 & J1733-1304 & J1924-2914 & J1820-2528 \\
    \hline
    [2016.1.00344.S] & S. P\'erez & 2017 Sep 18 & Ext 1 & 43.9 & 46 & 20 - 12000 & 231.985, 1.875 & J1733-1304 & J1924-2914 & J1826-2924 \\
    & & & & & & & 217.985, 1.875 & J1733-1304 & J1924-2914 & J1826-2924 \\
    & & 2017 Sep 19 & Ext 2 & 43.8 & 42 & 20 - 11500 & 231.985, 1.875 & J1924-2914 & J1924-2914 & J1826-2924 \\
    & & & & & & & 217.985, 1.875 & J1924-2914 & J1924-2914 & J1826-2924 \\
    & & 2017 Nov 9 & Ext 3 & 43.9 & 46 & 100 - 13500 & 231.990, 1.875 & J1924-2914 & J1924-2914 & J1826-2924 \\
    & & & & & & & 217.990, 1.875 & J1924-2914 & J1924-2914 & J1826-2924 \\
    & & 2016 Oct 4 & Com 1 & 26.8 & 40 & 17.5 - 2660 & 231.985, 1.875 & J1924-2914 & J1924-2914 & J1820-2528 \\
    & & & & & & & 217.985, 1.875 & J1924-2914 & J1924-2914 & J1820-2528 \\
    & & 2017 July 5 & Com 2 & 26.8 & 44 & 15 - 2630 & 232.004, 1.875 & J1924-2914 & J1924-2914 & J1826-2924 \\
    & & & & & & & 218.004, 1.875 & J1924-2914 & J1924-2914 & J1826-2924\\
    \hline
    \multicolumn{4}{c}{\scriptsize Ext: extended baseline configuration, Com: compact baseline configuration} \\
    \end{tabular}}
    \label{tab:almaobs}
\end{table*}

ALMA band 6 observations of the disc around HD\,169142 from Projects [2015.1.00490.S] and [2016.1.00344.S] (hereafter P2015 and P2016) were used to image the 1.3mm thermal continuum emission and the $^{12}$CO, $^{13}$CO and C$^{18}$O $J$=2-1 rotational line transitions. For specifications of the individual executions per project, refer to Table \ref{tab:almaobs}. We calibrated P2015 with the ALMA pipeline using {\sc casa} version 4.7.0, while P2016 was pipeline calibrated by the ESO (European Southern Observatory) staff. Prior to self-calibration, visibilities per execution were independently imaged using only the continuum spectral window to determine parameters for phase-centering prior to merging executions with shared antenna configurations. We used the publicly available {\sf reduction-utils} script by the DSHARP team to (1) determine parameters for phase-centering, which does so by fitting a Gaussian to the inner part of the disc, and (2) determine whether further flux-rescaling was required post-pipeline calibration. We re-scaled executions 1, 2 and 3 from P2015 due to an initial difference of >4$\%$ in flux. Finally, prior to self-calibration, we generated three concatenated files with (i) the three individual execution blocks of P2015, (ii) the two executions of P2016 with compact configuration and (iii) the three executions of P2016 with extended configuration, using {\sc casa} task {\sf concat}.

Self-calibration involved the sequential application of multiple phase-only calibrations to the continuum-only spectral windows (spws), with solution intervals in the order of [length of a scan] > 120s > 60s > 30s > 15s > 6s. An additional, amplitude+phase calibration was applied post phase-only calibration with a solution interval equal to the length of an entire scan. The compact (Com) and extended (Ext) configurations from P2016 and singular configuration from P2015 were independently self-calibrated. Following each iteration in self-calibration, the visibilities were imaged to ensure an increase in the peak signal-to-noise ratio (SNR). Continuum peak SNR improved by a factor of $\sim$4 and $\sim$6 for P2016 and P2015, respectively. The self-calibration tables were then applied to the corresponding line emission spws. For P2016, calibration tables for continuum spw centred at $\sim$231 GHz were used to self-calibrate the $^{12}$CO $J$=2-1 transition, whilst those for continuum spw centred at $\sim$217 GHz were used on $^{13}$CO and C$^{18}$O. For P2015, calibration tables from the single continuum only spw were applied to all line spws. P2015 has a channel spacing of 61kHz for the line spws, corresponding to a spectral resolution of 71kHz, due to the correlator applying Hanning smoothing. Similarly, P2016 has a channel width of 122kHz for line spws with a channel binning of 4. The product images are the combined products of the individually self-calibrated P2015 and P2016 data sets, where {\sc casa} tasks {\sf fixvis} and {\sf fixplanets} were used for alignment prior to merging using {\sc casa} task {\sf concat}.

Imaging was performed using {\sc casa} task {\sf tclean} with the {\sf multi-scale deconvolver} and several weighting schemes ({\sf super-uniform and Briggs with a few robust parameters}). In a few instances, a {\sf uv-taper} was added in order to smooth emission and circularise the synthesized (clean) beam. Masking was performed using {\sf auto-multithresh}. Continuum images produced using super-uniform weighting, Briggs r=0 and Briggs r=0 + uv-taper have beams of 0\farcs028$\times$0\farcs016, 0\farcs044$\times$0\farcs028 and 0\farcs065$\times$0\farcs059 with an RMS of 0.027\,mJy/beam, 0.011\,mJy/beam and 0.010\,mJy/beam, respectively. We images the $^{12}$CO, $^{13}$CO and C$^{18}$O $J$=2-1 transitions using Briggs weighting with r=0 and a uv-taper resulting in beams of 0\farcs073$\times$0\farcs067, 0\farcs069$\times$0\farcs063 and 0\farcs070$\times$0\farcs064 with an RMS of 1.58\,mJy/beam, 1.39\,mJy/beam and 1.06\,mJy/beam, respectively.

\begin{figure}
    \centering
    \includegraphics[width=0.49\textwidth]{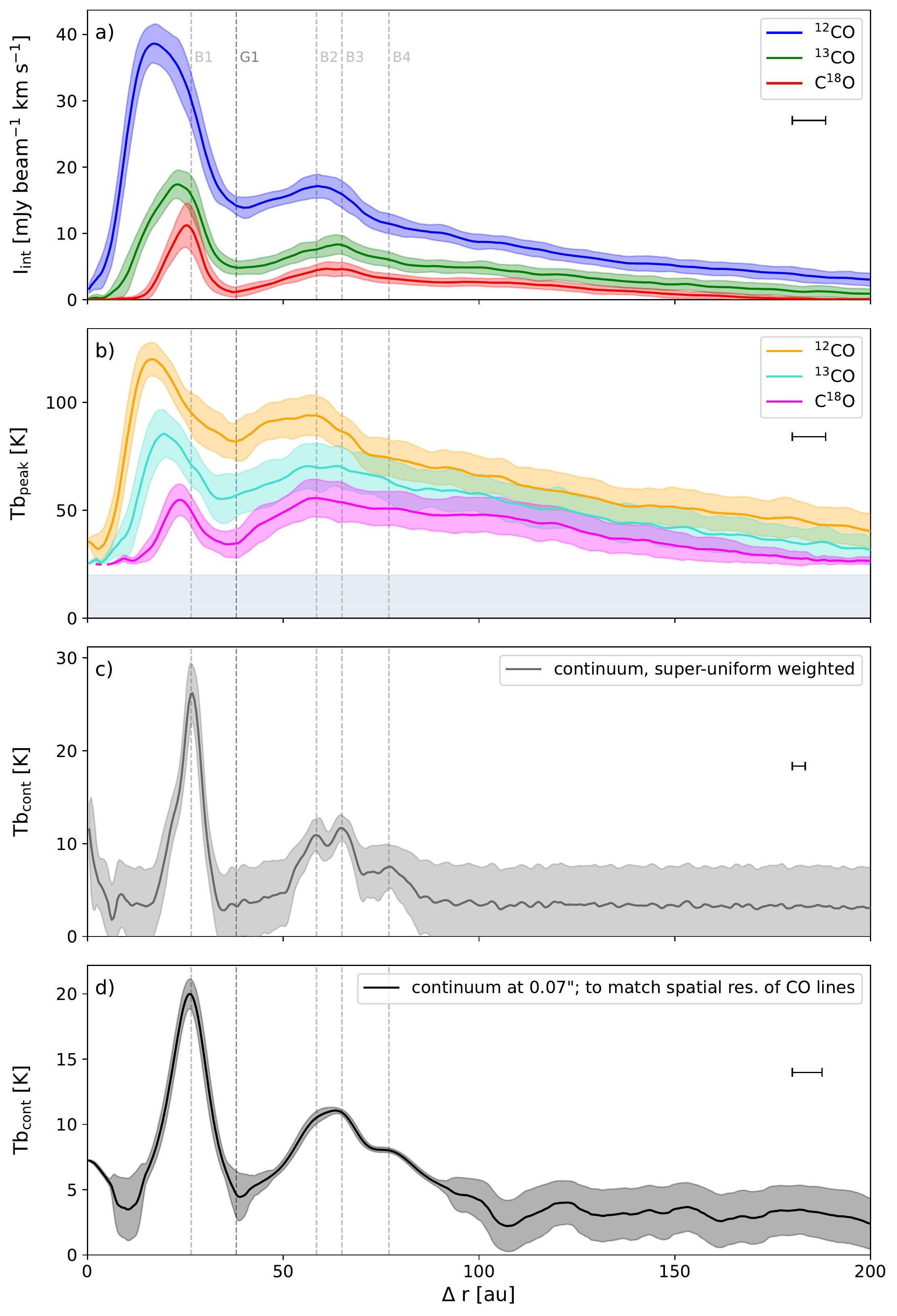}
    \caption{\textbf{Azimuthally averaged profiles}. a) non-continuum subtracted integrated intensities of the $^{12}$CO, $^{13}$CO and C$^{18}$O $J$=2-1 lines; b) non-continuum subtracted peak intensity maps of the $^{12}$CO, $^{13}$CO and C$^{18}$O $J$=2-1 lines, portrayed in units of brightness temperature. The blue highlighted region marks the freeze-out temperature of CO; c) super-uniform weighted 1.3mm continuum emission; d) 1.3mm continuum emission imaged for a resolution of 0\farcs07, to match spatial resolution of line emission. The shaded regions represent the 1$\sigma$ uncertainty for the corresponding line colour. The light grey vertical lines mark the locations of the dust rings (same labels as in \citealp{Perez2019}), and the dark grey vertical line marks the location of the gap between the CO rings (G1). The horizontal bar represents the spatial resolution.}
    \label{fig:COprofiles}
\end{figure}

\section{Analysis methods}

\subsection{Searching for kinematic signatures of companions}

\label{sec:kinematic_sigs}
For gas flow along circular orbits, the rotational velocity of the gas is given by \citep[e.g.][]{Rosenfeld2013}
\begin{equation}
    \frac{v_{\mathrm{gas}}^2}{r} = \underbrace{\frac{GM_{*}r}{(r^{2}+z^{2})^{3/2}}}_{1} + \underbrace{\frac{1}{\rho_{\mathrm{gas}}} \frac{\partial P_{\mathrm{gas}}}{\partial r}}_{2} + \underbrace{\frac{\partial \phi_{\mathrm{gas}}}{\partial r}}_{3},
    \label{eq:vrot}
\end{equation}
where (1) is the Keplerian velocity component for a geometrically thick disc, where $r$ is the cylindrical radius and $z$ is the height above the disc midplane. (2) is the radial pressure gradient component (where P$_{\mathrm{gas}}$ = $\rho_{\mathrm{N}}$k$_{\mathrm{B}}$T), and (3) is the self-gravity component, where $\phi_{\mathrm{gas}}$ is the gravitational potential of the disc; generally only significant when M$_{\mathrm{disc}}$/M$_{*}$ $>$ H/R. For a typical disc, globally both density and temperature decrease as a function of radius which translates into $\partial P_{\mathrm{gas}}/{\partial r} < 0$, and thus, a global sub-Keplerian profile.

To search for hidden kinematic signatures of an embedded planet(s) in HD\,169142, we utilise the software {\sc eddy} \citep{eddy}, which subtracts a background disc model with parameters determined via the Markov chain Monte Carlo (MCMC) method, where the posteriors explore the azimuthally averaged velocity fields from the observations. We fit for disc centre coordinates (x$_{0}$ and y$_{0}$), PA, \msun, and v$_{\mathrm{LSR}}$ using 250 walkers to explore the posterior distribution with a total of 10000 steps, of which the walkers were found to have converged in less than 500 steps. Disc inclination was kept fixed at 13$^{\circ}$. The MCMC fitting was limited to the inner 1\farcs5 radial range of the disc, because (1) line emission past this radius was found to be more noise dominated, and (2) the gas surface density was also found to drop drastically past the outer-most dust ring and thus could have resulted in an underestimation of the stellar mass if included. Additionally, we also mask the inner 0\farcs2 (2\,$\times$\,beam width) to exclude kinematic biasing from beam smearing. A summary of the best-fit parameters for all three tracers are listed in Table \ref{tab:mcmcfitparams}. We refrain from including the associated uncertainties, as the MCMC fitting produces statistically negligible uncertainties for the free parameters. As an alternative, we produce a series of residual velocity maps to assess the significance of variations in PA, \msun, and inclination for $^{12}$CO (Appendix \ref{sec:eddy_uncertainty}).

From the computed gas surface density profile (Fig.\ref{fig:col_den}) it is evident that past 80\,au (0\farcs68) the gas rotational profile should be sub-Keplerian due to the negative density gradient. Therefore, in {\sc eddy} we also fit for a tapered Keplerian background model following the methodology outlined in \cite{Teague2022}. A tapered Keplerian model follows the form,
\begin{equation}
    v_{\mathrm{kep, tapered}} = \sqrt{\frac{G(M_{*}+M_{\mathrm{d}}(r))}{r}} ,
\end{equation}
where $M_d(r)$ is an effective disk mass set by
\begin{equation}
    M_{d}(r) = M_{\mathrm{disk}} \times \frac{r^{2-\gamma} - r^{2-\gamma}_{\mathrm{in}}}{r^{2-\gamma}_{\mathrm{out}} - r^{2-\gamma}_{\mathrm{in}}} .
\end{equation}
r$_{\mathrm{in}}$ and r$_{\mathrm{out}}$ define the inner and outer boundaries where gas rotational velocity deviates from Keplerian, respectively. M$_{\mathrm{disk}}$ is measured at r$_{\mathrm{out}}$ (set to 3\farcs0 in our case) and a negative quantity translates to slowed rotation, mimicking a pressure gradient, whilst $\gamma$ determines how quickly the gas rotational profile slews from Keplerian.

Here we include M$_{\mathrm{disk}}$, $\gamma$, r$_{\mathrm{in}}$ as additional free parameters to the MCMC fitting and provide the best fit values in Table
\ref{tab:mcmcfitparams} for each tracer. To determine the statistical significance of any detected kinematic structure, we divide the residual velocity maps by the statistical uncertainty on the velocity per pixel ($\delta$v$_{\mathrm{0}}$).

\begin{figure*}
    \centering
    \includegraphics[width=\textwidth]{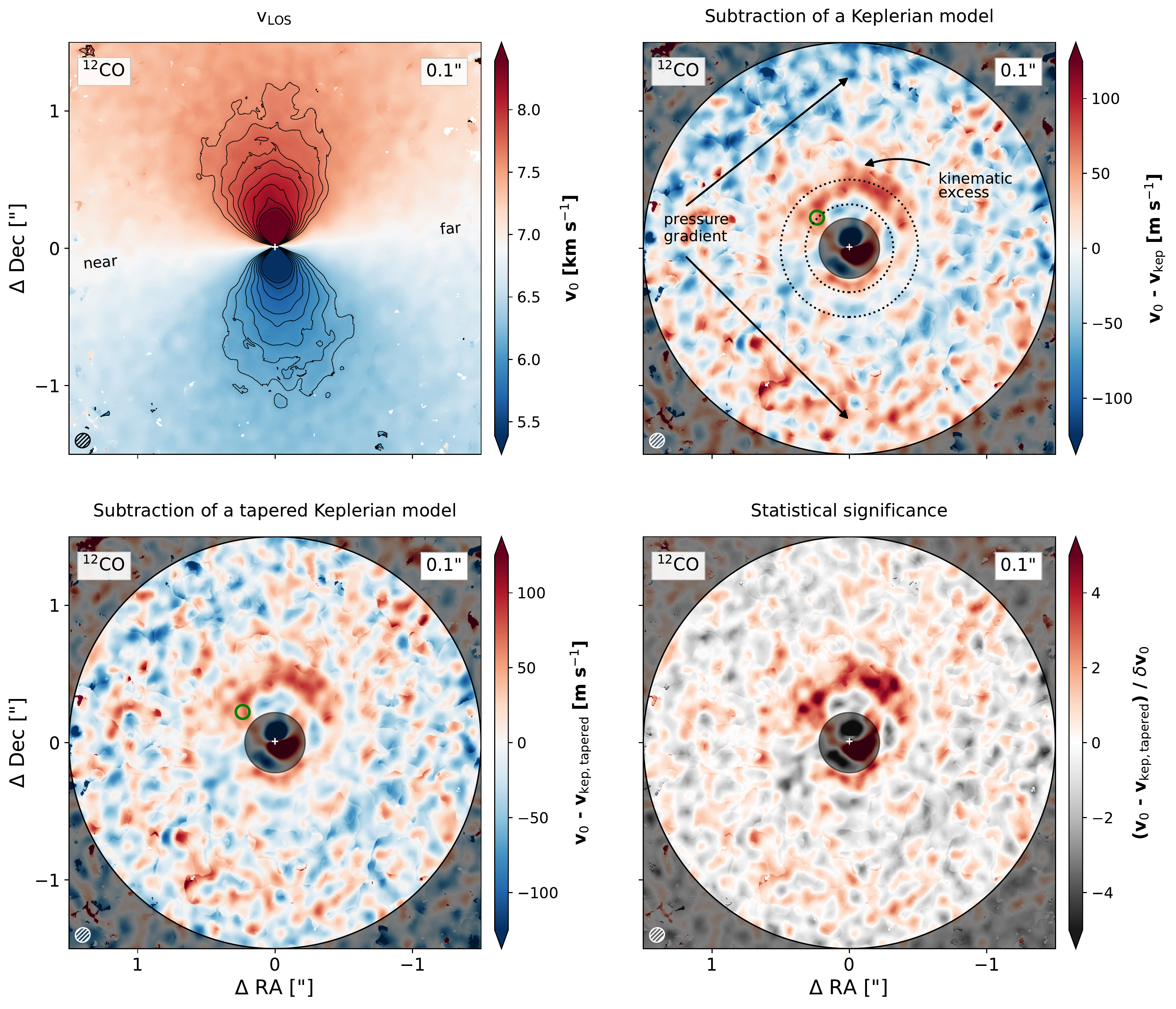}
    \caption{\textbf{Kinematic excess.} Top left: line-of-sight velocity map of $^{12}$CO, at a resolution of 0\farcs1. Top right: residual velocity maps post-subtraction of a Keplerian disc model. Bottom left: residual velocity maps post-subtraction of a tapered Keplerian model. Bottom right: residual velocities in panel three divided by the statistical uncertainty in the line-of-sight velocity maps. The grey shaded area represents the masked-out regions in the MCMC fitting.The dotted rings represent the G1[38au] gap and B2[59au] dust ring, overlaid for comparison. The green circle marks the location of the high intensity point source (blob D) reported in \citet{Gratton2019}.}
    \label{fig:gv0+eddy_map_12co}
\end{figure*}

\begin{figure*}
    \centering
    \includegraphics[width=\textwidth]{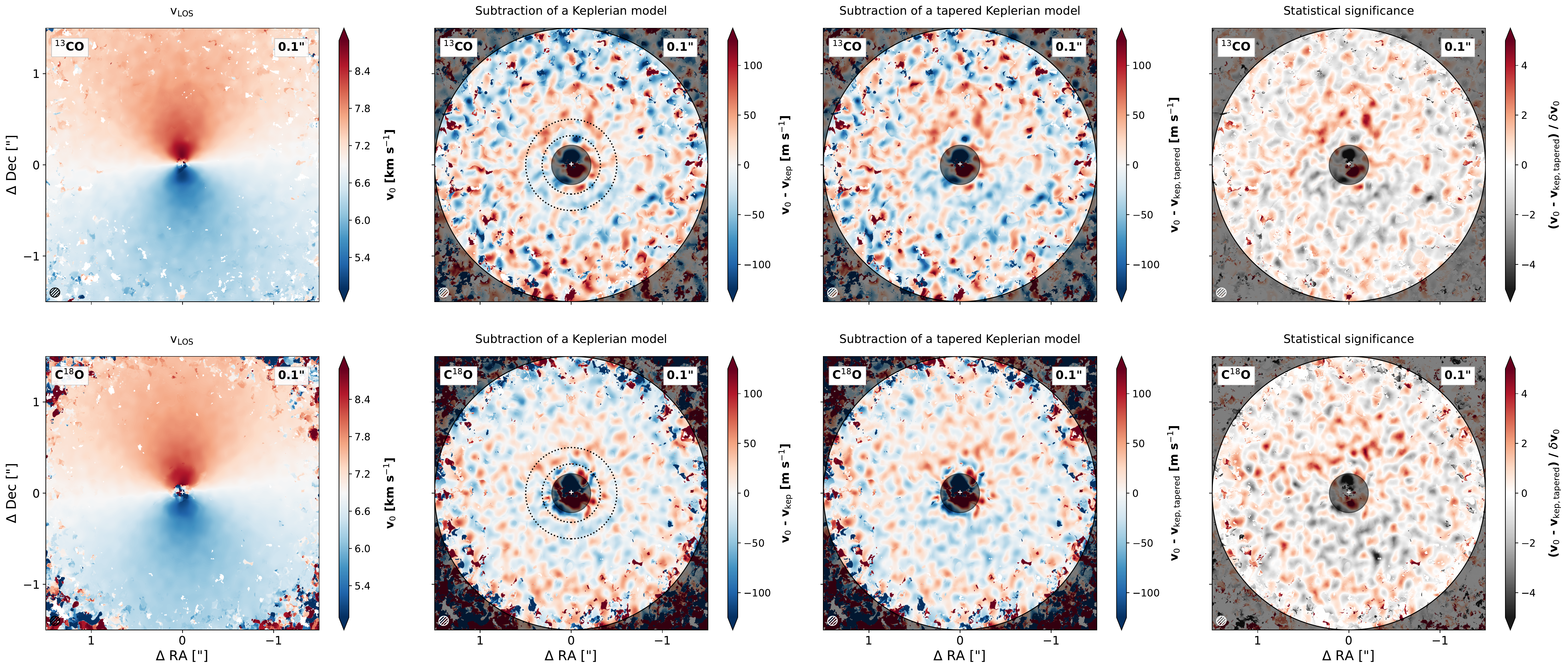}
    \caption{\textbf{Same as Figure \ref{fig:gv0+eddy_map_12co} but for $^{13}$CO and C$^{18}$O.} Top panel: $^{13}$CO. Bottom panel: C$^{18}$O. First column: line-of-sight velocity map. Second column: residual velocity maps post-subtraction of a Keplerian disc model. Third column: residual velocity maps post-subtraction of a tapered Keplerian model. Fourth column: residual velocities in column 3 divided by the statistical uncertainty in the line-of-sight velocity maps. The grey shaded area represents the masked-out regions in the MCMC fitting. The dotted rings represent the G1[38au] gap and B2[59au] dust ring, overlaid for comparison. The green circle marks the location of the high intensity point source (blob D) reported in \citet{Gratton2019}.}
    \label{fig:gv0+eddy_maps_13co_c18o}
\end{figure*}

To spectrally collapse line emission (refer to Figs. \ref{fig:gv0+eddy_map_12co} and \ref{fig:gv0+eddy_maps_13co_c18o}), we utilise the {\sc bettermoments} package \citep{bettermoments}, which returns both the projected peak velocity maps (v$_{\phi}$) and the statistical uncertainty per pixel ($\delta$v$_{\phi}$). Specifically, we use the {\sf Gaussian} method to spectrally collapse the cube, as the method has been shown to have minimal statistical uncertainty \citep{Yu2021}.

For the kinematic analysis we use maps with a slightly degraded beam of 0\farcs1 resolution, using the Briggs robustness factor of 0.5 (instead of 0) to achieve higher sensitivity. Additionally, a uv-taper is also applied to circularise the beam. Due to the low inclination of the disc, the background disc model is assumed to be geometrically thin \emph{e.g.} no vertical extension (z=0). Theoretically, this assumption should result in an underestimation of the true stellar mass (in accordance with Equation \ref{eq:vrot}) for the $^{12}$CO and $^{13}$CO tracers which are typically more elevated in the disc layers. However, the best-fit stellar masses of 1.473, 1.488 and 1.487 \msun\, for $^{12}$CO, $^{13}$CO and C$^{18}$O respectively, are found to only differ by $\sim$1\%. To extract the three independent velocity profiles (v$_{\phi}$, v$_{\mathrm{r}}$, and v$_{\mathrm{z}}$), we use the {\sf fit annuli} function in eddy on the line-of-sight velocity maps computed with the {\sf Gaussian} method.

\subsection{Computing the gas surface density structure}
\label{sec:gas_surface_den}

\begin{figure}
    \centering
    \includegraphics[width=0.45\textwidth]{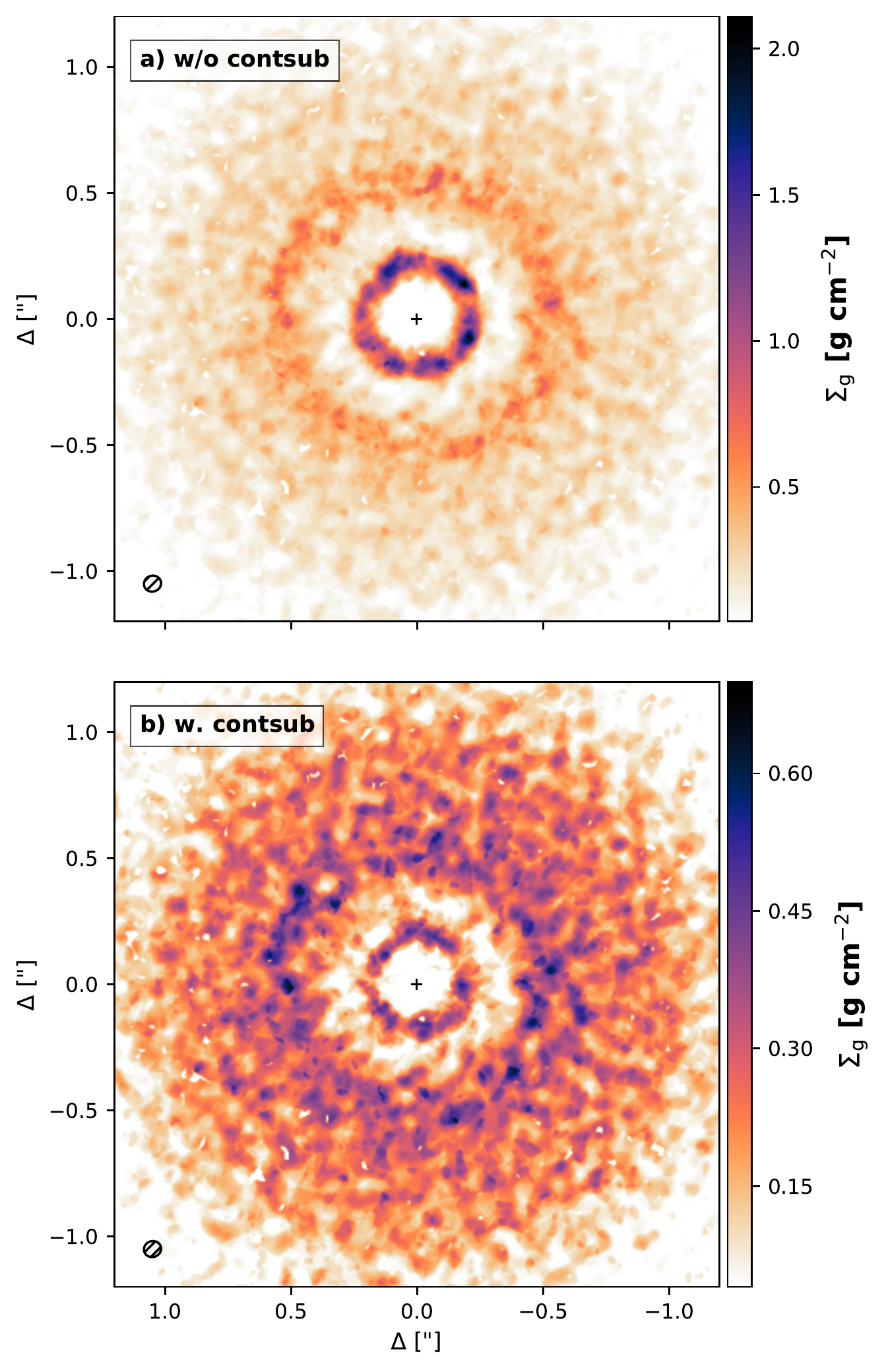}
    \caption{\textbf{Gas column density maps} computed for when excluding (a) and including (b) the effects of continuum subtraction on line emission. Beam is presented in the bottom-left corner. Both maps are deprojected for a disc inclination of 13$^{\circ}$ along a PA of 95$^{\circ}$.}
    \label{fig:col_den}
\end{figure}

The upper and lower bounds of the gas column density structure are computed as per the method outlined in \cite{Garg2021}, and we refer the reader there for more thorough discussions on caveats and a means of accounting for them. This technique has also been implemented in \citet{Lyo2011, Schwarz2016, Perez2015, Casassus2021}. For completeness, we provide the main formulae in Appendix \ref{sec:appendix1}. We use the Leiden Atomic and Molecular Database (LAMDA) \citep{Schoier2005} (accessed in July 2020), to acquire quantities for the $E_u$ and $A_{ul}$ coefficients in Equations \ref{eq:col1} and \ref{eq:col2}. The computed molecular column density using Equations \ref{eq:col1} and \ref{eq:col2} is scaled to represent total gas column density by multiplying by the canonical ISM abundance ratios: [$^{12}$C]/[$^{13}$C] $\approx$ 70 \citep{Stahl2008}, [$^{16}$O]/[$^{18}$O] $\approx$ 500 \citep{Wilson1994} and [H$_{2}$]/[$^{12}$CO] $\approx$ 10$^{4}$. The total gas column density structure calculated considering either continuum subtracted or non-continuum subtracted line maps is presented in Figure \ref{fig:col_den}. We made no assumption on the level of the individual isotopic depletion in the disc, as the ambiguity in the true line intensity can already lead to an order of magnitude uncertainty on the computed column density profile, \emph{i.e.} the difference between the upper and lower bounds due to continuum subtraction. Additionally, this method of computing the surface density requires that the temperature and density tracing species share the same local temperature. \cite{Pinte2018} demonstrated that the emission heights of the $^{13}$CO and C$^{18}$O $J$=2-1 surfaces closely overlap, therefore given that we find $^{13}$CO to be optically thick throughout the disc, we used it as the temperature tracer (rather than $^{12}$CO), whilst we used the optically thin C$^{18}$O to derive the column density following Appendix \ref{app:CD}.

\begin{figure*}
    \centering
    \includegraphics[width=\textwidth]{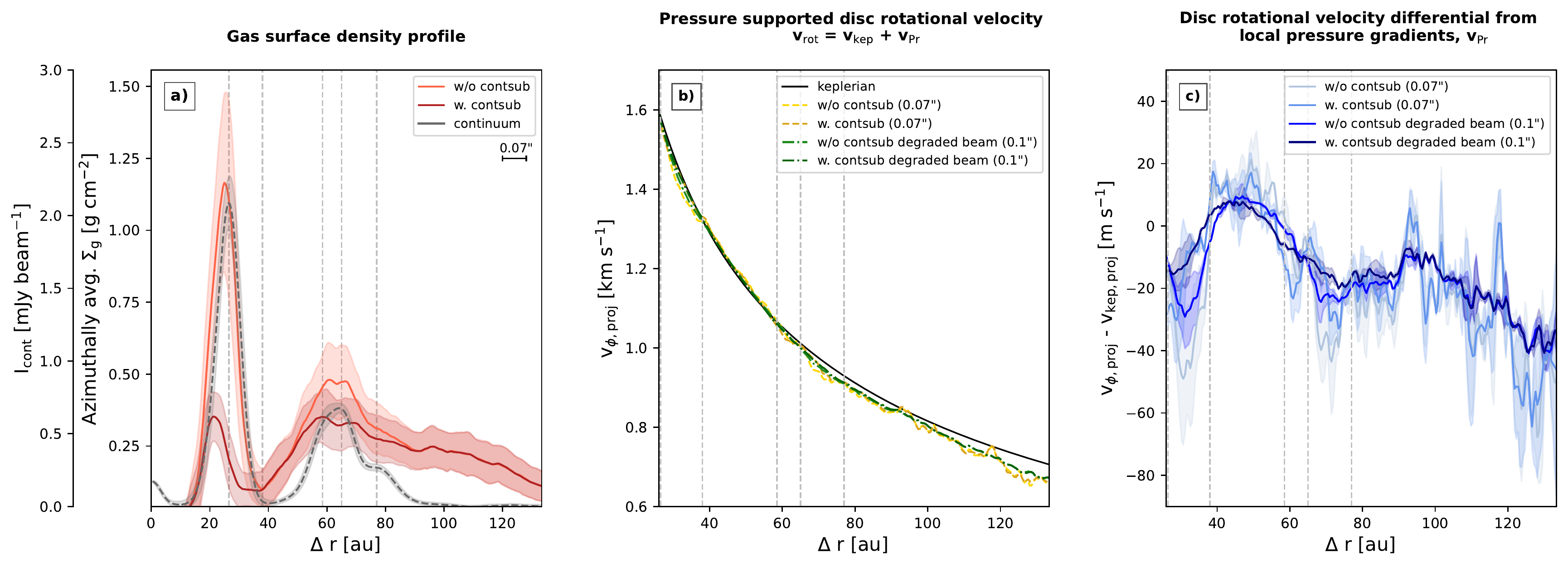}
    \caption{\textbf{Rotational velocity profile of a pressure supported disc}. a) azimuthally averaged gas surface density profiles of the 2D maps shown in Figure \ref{fig:col_den}. The figure is overlaid with the azimuthally averaged continuum emission at 0\farcs07 resolution shown in Figure \ref{fig:COprofiles}\,d). b) presents gas rotational velocity for a pressure supported disc, using the azimuthally averaged gas density profile in panel a. Here we use both the continuum subtracted and non-continuum subtracted density profiles. c) expected gas rotational velocity differential due to the local pressure gradients. The shaded regions represent the 1$\sigma$ uncertainty. Additionally, for panels b and c we also compute the profiles for a more degraded beam 0\farcs1 (the resolution of the line-of-sight velocity maps in Figs. \ref{fig:gv0+eddy_map_12co} and \ref{fig:gv0+eddy_maps_13co_c18o}, and overlay these on the plots. The dashed vertical lines represent the  G1[38au] gap and B1[26au], B2[59au], B3[66au] and B4[77au] dust rings.}
    \label{fig:pvp}
\end{figure*}

From the gas column density and temperature maps, we are able to compute the gas rotational profile for a fully pressure supported disc as well as deduce radial and azimuthal velocity offsets expected from localised pressure gradients. Given that the disc's gas structure is fairly symmetric, we use an azimuthally averaged gas surface density profile to compute the second term in Equation \ref{eq:vrot} and present the results in Figure \ref{fig:pvp}.

The gas volume density is computed via
\begin{equation}
    n(r)_{\mathrm{gas}} = \frac{\Sigma(r)}{\sqrt{2\pi}h(r)},
\end{equation}
where $\Sigma$(r) is the gas surface density profile and $h(r)$ is the hydrostatic scale height given by
\begin{equation}
    h(r) = \sqrt{\frac{r^{3}k_{B}T(r)}{GM_{*}\mu m_{H}}},
    \label{eq:h}
\end{equation}
under the assumption of an isothermal vertical temperature structure. $k_{\mathrm{B}}$ is the Boltzmann constant, $G$ is the gravitational constant, M$_{*}$ is the stellar mass, $\mu$ is the mean molecular mass ($\sim$2.3), and $m_{\mathrm{H}}$ is the mass of hydrogen.

Pressure is given by
\begin{equation}
    P(r) = \rho_{N, \mathrm{gas}}(r) k_{B}T(r),
\end{equation}
where T(r) is equivalent to the azimuthally averaged brightness temperature profile of $^{13}$CO and $\rho_{N, \mathrm{gas}}(r)$ is the gas number density.

%--------------------------------------------------
\begin{table}
    \caption{Best-fit parameters from MCMC fitting of the velocity fields in {\sc eddy}.}
    \setlength{\tabcolsep}{5pt}
    \resizebox{0.45\textwidth}{!}{%
    \begin{tabular}{lcccccc}
    \multicolumn{2}{l}{\textbf{Keplerian model}}\\
    \hline
    & \textbf{PA} ($^{\circ}$) & \textbf{M$_{*}$} (\msun) & \textbf{v$_{\mathrm{LSR}}$} (ms$^{-1}$) & & &\\
    \hline
    \textbf{$^{12}$CO} & 5.33 & 1.47 & 6897 & & &\\[2mm]
    \textbf{$^{13}$CO} & 4.96 & 1.49 & 6895 & & &\\[2mm]
    \textbf{C$^{18}$O} & 5.15 & 1.49 & 6886 & & &\\[2mm]
    \hline\\
    \multicolumn{3}{l}{\textbf{Tapered Keplerian model}}\\
    \hline
    & \textbf{PA} ($^{\circ}$) & \textbf{M$_{*}$} (\msun) & \textbf{v$_{\mathrm{LSR}}$} (ms$^{-1}$) & \textbf{M$_{\mathrm{disk}}$} (\msun) & \textbf{$\gamma$} & \textbf{r$_{\mathrm{in}}$} (\farcs)\\
    \hline
    \textbf{$^{12}$CO} & 5.32 & 1.47 & 6895 & -0.59 & 0.37 & 0.70\\[2mm]
    \textbf{$^{13}$CO} & 4.95 & 1.49 & 6894 & -0.59 & -0.27 & 0.74\\[2mm]
    \textbf{C$^{18}$O} & 5.17 & 1.49 & 6886 & -0.59 & -0.74 & 0.74\\[2mm]
    \hline
    \multicolumn{4}{l}{\scriptsize}\\
    \end{tabular}}
    \label{tab:mcmcfitparams}
\end{table}

\begin{figure*}
    \centering
    \includegraphics[width=0.9\textwidth]{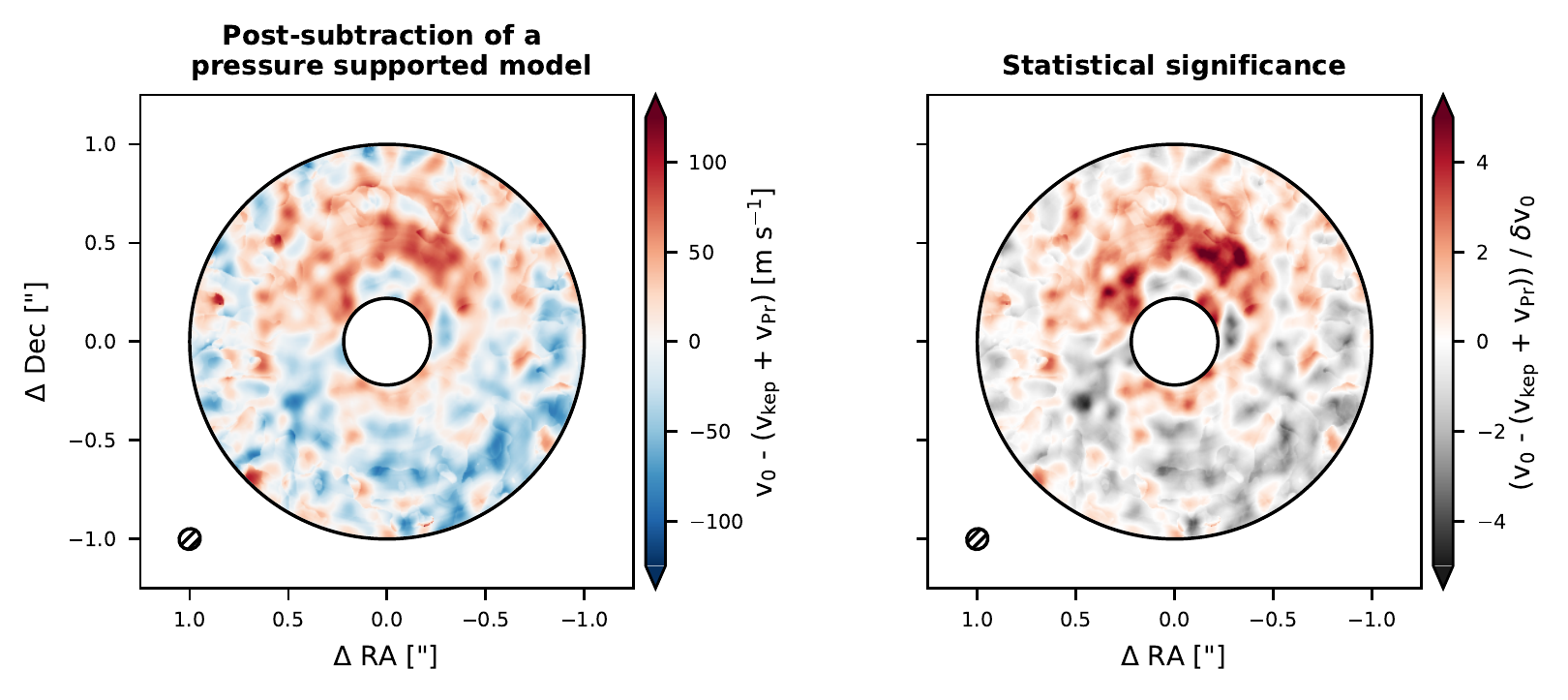}
    \caption{\textbf{Velocity residual post-subtraction of a fully pressure supported disc}. Left: $^{12}$CO velocity residuals map post-subtraction of pressure supported gas rotation. Right: residual velocities in the left panel divided by the statistical uncertainty in the line-of-sight velocity map. Here we apply an outer mask at 1\farcs0 due to the gas surface density profile becoming noise dominated past this radii. The dotted rings represent the G1[38au] gap and B2[59au] ring. The green circle marks the location of blob D reported in \citet{Gratton2019}. Beam is shown on the bottom left corner.}
    \label{fig:vpr}
\end{figure*}

\subsection{Semi-analytical modelling of the wake}

For qualitative comparison of the kinematic features seen in the observations to those induced by an embedded planet, we employ the semi-analytical method of modelling the wake presented in \cite{Bollati2021}, with the implementation in {\sc wakeflow}\footnote{\url{https://github.com/TomHilder/wakeflow}} . Via this method, we only consider co-planar orbits and velocity deviations in the plane of the disc, thus neglecting vertical motions. We defer comparisons with full 3D hydrodynamical models for deeper ALMA observations. The semi-analytical model assumes power-law functions for surface density ($\Sigma$) and sound speed (c) profiles, given by,
\begin{equation}
    \Sigma_{0}(\mathrm{r}) = \Sigma_{\mathrm{p}}(r/r_{\mathrm{p}})^{-\delta},
\end{equation}
and
\begin{equation}
    c_{0}(\mathrm{r}) = c_{\mathrm{p}}(r/r_{\mathrm{p}})^{-q},
\end{equation}
respectively. Subscript $p$ denotes estimates at the location of the planet. Here we use 1.0 and 0.2 for the $\delta$ and $q$ power law indexes; average values derived from fitting to the surface density and brightness temperature profile. The shape of the wake ($\varphi$) is inversely proportional to the disc aspect ratio (\emph{e.g.} $\varphi$ $\propto$ (h$_{\mathrm{p}}$/r$_{\mathrm{p}}$)$^{-1}$), which we determine and set to 0.08 (refer to Section \ref{sec:mp_est}). Models are generated for planetary masses of 1\,M$_{\mathrm{J}}$ and 10\,M$_{\mathrm{J}}$, calculated for disc viscosities of 10$^{-3}$ and 10$^{-1}$, respectively (refer to Sections \ref{sec:mp_est} and \ref{sec:discussion_planetmass} for a more thorough discussion). The planet is placed at an orbital distance of 38\,au and PA of 43.8$^{\circ}$, as per the location of blob D in \cite{Gratton2019}.

The surface density and velocity field calculated by {\sc wakeflow} are passed to the radiative transfer code {\sc mcfost} \citep{Pinte2006mcfost, Pinte2009mcfost}. The surface density is extended vertically assuming a Gaussian profile with $h(r) = h_0 (r/r_0)^{1.125}$, where we set $h_0 = 2.95$au at $r_0 = 38$au. We assume that the velocity fields $v_r$ and $v_\phi$ do not depend on the altitude in the disc. We set the total gas mass to $10^{-2}$\,M$_\odot$\citep{Toci2020} and the gas-to-dust ratio (which is constant across the disc) to 100. We assume a grain size distribution $\mathrm{d}n(a) \propto a^-3.5\mathrm{d}a$ between 0.03$\mu$m and 1mm, with a silicate composition \citep{Weingartner01}. Dust properties are calculated using the Mie theory. We assume that the disc is passively heated by the star, and use 12.8 millions packets to compute the dust temperature structure. The generated synthetic $^{12}$CO maps have a channel spacing of 32\,ms$^{-1}$, assuming the line is in LTE with T$_\mathrm{gas}$ = T$_\mathrm{dust}$. We use a relative CO abundance of $10^{-4}$ and include CO freeze-out below 20K, photodissociation, photodesorption (following \citealp{Pinte2018}). The synthetic cubes are then convolved both spatially (with the observed beam of 0.1") and spectrally (with a Hanning function of width 167\,ms$^{-1}$). To produce the residual velocity maps from the analytic models, we produce synthetic cubes with {\sc mcfost} identical to the setup above with the exception of the analytic velocity perturbations from {\sc wakeflow}, which are then subtracted from the {\sc wakeflow} + {\sc mcfost} cubes.

%--------------------------------------------------
\begin{figure*}
    \centering
    \includegraphics[width=\textwidth]{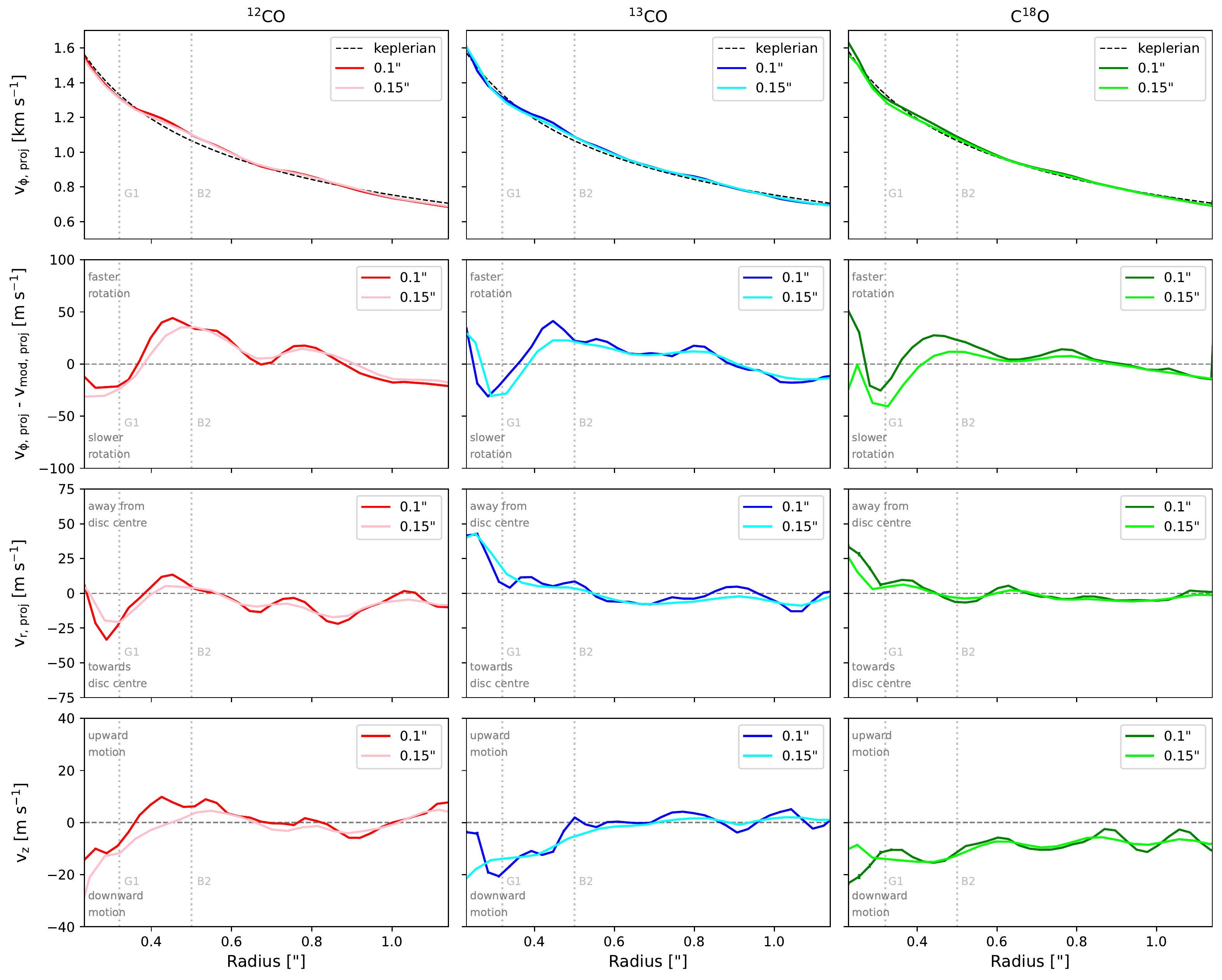}
    \caption{\textbf{Velocity components.} Row 1: azimuthally averaged gas rotational velocity v$_{\phi}$ for each tracer. Rows 2-4: profiles of the component velocities extracted from the velocity residual maps shown in Figures \ref{fig:gv0+eddy_map_12co} and \ref{fig:gv0+eddy_maps_13co_c18o}: rotational (v$_{\phi}$ - v$_{\mathrm{mod}}$), radial (v$_{\mathrm{rad}}$), and vertical (v$_{\mathrm{z}}$), respectively. For the Keplerian profiles a stellar mass of 1.48\msun\,is used. Velocity profiles for a further degraded beam of 0\farcs15 are also overlaid.}
    \label{fig:eddy_v_profiles}
\end{figure*}

\section{Results}

\subsection{Central cavity and rings in line emission}
\label{sec:COrings}

Figure \ref{fig:COmaps,intensity} displays the integrated intensity maps with and without continuum subtraction, and the peak intensity (I$_{\mathrm{peak}}$; portrayed in brightness temperature) maps without continuum subtraction, for the $^{12}$CO, $^{13}$CO and C$^{18}$O lines. A prominent gas ring (hereafter: R1) encompassing the central cavity is readily observed with all three tracers, along with a second (hereafter: R2) but relatively fainter and diffuse ring further out.

Figure \ref{fig:dustmaps} displays the 1.3mm thermal continuum emission first presented in \cite{Perez2019}, but as self-calibrated and imaged by us, and shown for several more weighting schemes to highlight the variation in dust structure when addressed at different resolutions throughout the paper.

For a quantitative analysis of the extent and widths of the gas rings and gaps/cavities, we deprojected the moment maps assuming an inclination of 13$^{\circ}$ perpendicular to the semi-major axis of the disc at PA=5$^{\circ}$ and applied azimuthal averaging. Profiles of the azimuthally averaged gas integrated flux and I$_{\mathrm{peak}}$ maps, with a comparison to the profiles of continuum emission imaged at super-uniform weighting and Briggs r=0.5 weighting with a uv-taper, where the latter more closely matches the resolution of the calibrated line emission, are presented in Figure \ref{fig:COprofiles}.

Gas emission towards the disc centre is heavily depleted with a sharp gradient in $^{12}$CO halting at $\sim$20\,au; the location of R1 in $^{12}$CO. Similar characteristics are also seen in the profiles of the optically thinner tracers, $^{13}$CO and C$^{18}$O, but with the peak emission of R1 shifting to larger radii of r$\approx$22 and r$\approx$26 au, respectively. Here, the peak of R1 in C$^{18}$O is almost consistent with the profile in continuum emission. The observed narrowing of the central gas cavity (R$_{\mathrm{cav}}$) from C$^{18}$O $\rightarrow$ $^{13}$CO $\rightarrow$ $^{12}$CO is due to an optical depth ($\tau$) effect, where a decreasing $\tau$ requires higher column densities before the molecular emitting layer becomes prominent. Hence, peak emission of R1 is seen to shift to an increasing radius from $^{12}$CO $\rightarrow$ $^{13}$CO $\rightarrow$ C$^{18}$O, but the subsequent drop in intensity occurs at a similar radius, of $\sim$40\,au, for all three tracers; the first gap (hereafter: G1). In comparison to R1, the emission from the second gas ring (R2) is found to be more diffuse and shallower in integrated intensity. In non-continuum subtracted I$_{\mathrm{peak}}$ maps, R2 in $^{12}$CO and $^{13}$CO is seen to be almost two-thirds as bright as R1, whilst in C$^{18}$O, the brightness temperature of R2 is comparable to R1. This disparity is likely due to the contrast in continuum emission between the two rings.

Given the inherent difference in optical depth between the tracers, the similarities seen between the intensity distributions of all three tracers is instead suggestive that the brightness profile of $^{12}$CO is not solely due to a local enhancement in temperature, instead an accumulation of gas in confined ring structures. Furthermore, the brightness temperature profiles of all three tracers remains above the CO freeze-out temperature (T$_{\mathrm{freeze}}\sim$20K) within the inner 200\,au radius, suggesting that the location of the rings seen in this source are not correlated to condensation fronts of carbon monoxide, or any other species with a freeze-out temperature less than 20K.

\subsection{A kinematic excess in the annular gap}

A kinematic excess azimuthally spanning a PA range of $\approx$-60 to 45$^{\circ}$ with a magnitude of $\sim$75ms$^{-1}$, is detected in the residual velocities of $^{12}$CO, post-subtraction of a Keplerian model (Fig.\ref{fig:gv0+eddy_map_12co}). Radially the feature extends between the B1 and B2 dust rings, and at PA of $\sim$43.8$^{\circ}$ is found to overlap well with the high-intensity residual (blob D) reported in \citet{Gratton2019} at a radius of $\sim$38\,au; represented as a green circle in Figure \ref{fig:gv0+eddy_map_12co}. Counterparts of this feature at lower disc altitudes ($^{13}$CO and C$^{18}$O residual velocity maps; Fig.\ref{fig:gv0+eddy_maps_13co_c18o}) are found to be near absent however. This may be due to either (1) insufficient integration time of the fainter tracers to resolve a low magnitude velocity deviation, or (2) a variation in the gas rotation profile between the three tracers such that any deviations in the vertical velocity component are near absent for the midplane tracers.

A break down of the $^{12}$CO velocity residuals into the v$_{\phi}$, v$_{\mathrm{r}}$, and the v$_{\mathrm{z}}$ components (Figure \ref{fig:eddy_v_profiles}) reveal velocity perturbations with magnitudes of $\approx$50\,ms$^{-1}$, 40\,ms$^{-1}$ and 20\,ms$^{-1}$, respectively. The extraction of the three velocity components is performed on azimuthally averaged velocity residuals, hence the magnitude of the perturbations are lower than the maximum deviations seen in the 2D velocity residual maps. We also extract velocity profiles for a further degraded beam of 0\farcs15 (overlaid in Fig. \ref{fig:eddy_v_profiles}), and find that the general profile of the velocity residuals at this resolution remains consistent with those extracted at the 0\farcs1 resolution, suggesting that sub-beam intensity gradients do not play a significant role in the extracted rotation curves.

\subsection{Background velocity deviations from the pressure gradient}

In {\sc eddy} we subtract both a Keplerian only model and a tapered Keplerian model where the latter better mimics the slowed gas rotational profile in the outer regions (>100au). However, from Figures \ref{fig:COmaps,intensity} and \ref{fig:dustmaps} it is evident that even interior to 100\,au the disc harbors multiple localised pressure gradients associated with the concentric rings seen in both dust and gas emission.

In Figure \ref{fig:pvp}, we use the azimuthally averaged gas surface density profile to compute the global gas rotational velocity profile for a pressure supported disc (for both the upper and lower bounds of the gas density profile). We show the velocity deviations introduced from the localised pressure gradients in the right panel. We also overlay these perturbations with those from a more degraded beam ($\sim$0\farcs1) to match the angular resolution of the line maps used in {\sc eddy}. For a 0\farcs1 beam, the kinematic deviation from the pressure gradient is approximately 10\,ms$^{-1}$ between the G1 gap and B2 dust ring. Including the self-gravity term from Equation \ref{eq:vrot} is found to have negligible effect on the background velocity perturbations (difference of $\sim$0.25\,ms$^{-1}$ at 100\,au).

In Figure \ref{fig:vpr}, we use the continuum subtracted azimuthally averaged pressure gradient profile along with a Keplerian profile to compute a fully pressure supported background disc model. This model is then subtracted from the line-of-sight $^{12}$CO map (Fig.\ref{fig:gv0+eddy_map_12co}). The shape of the arc is found to be identical between the residual velocity maps post-subtraction of a Keplerian-only model and a pressure supported gas rotational velocity model. The magnitude of the offset is however found to be slightly lower at $\sim$60-70\,ms$^{-1}$. This demonstrates that the $^{12}$CO kinematic arc is in excess of deviations introduced from local pressure gradients in the disc.

\begin{figure*}
    \centering
    \includegraphics[width=\textwidth]{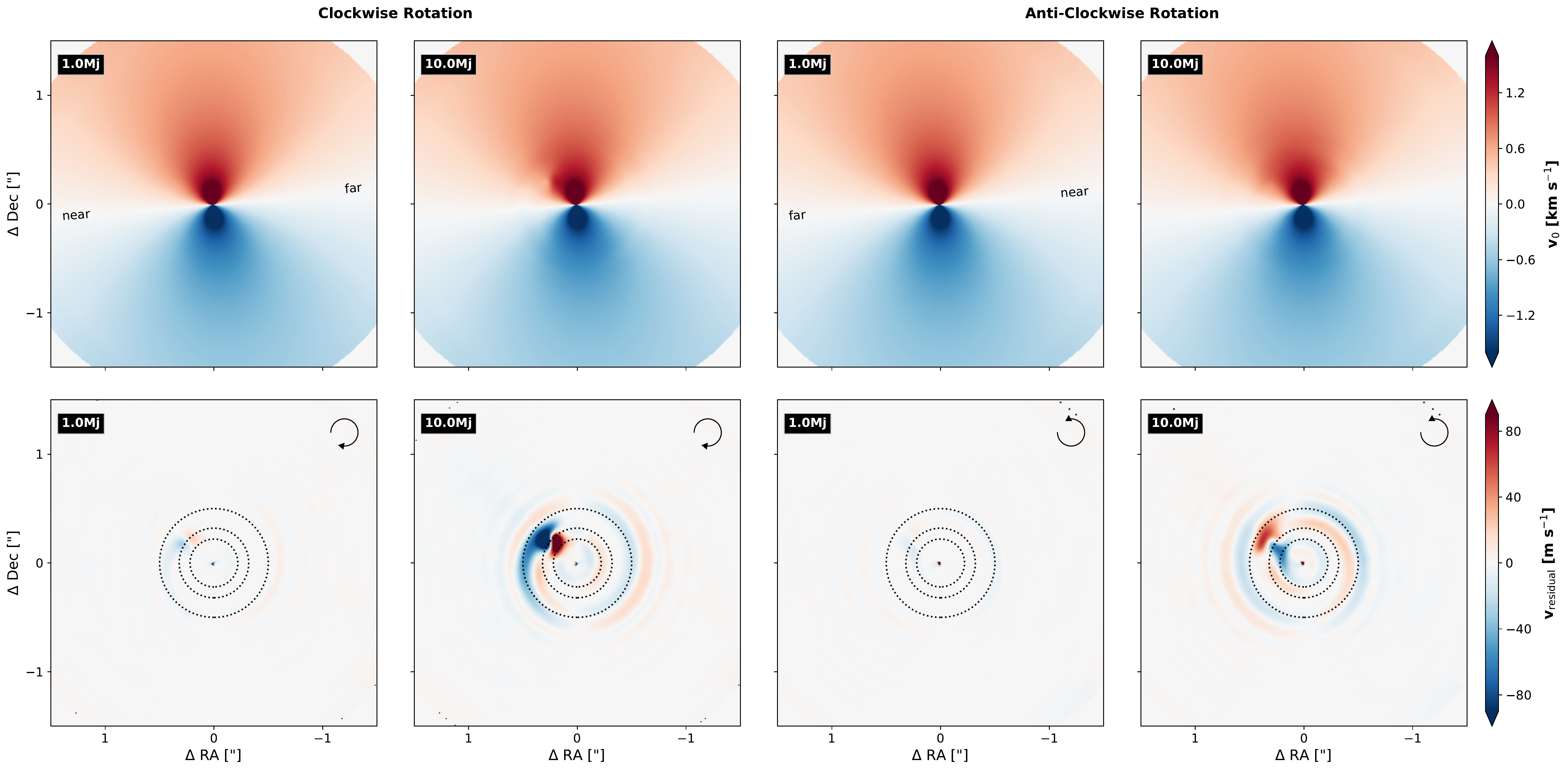}
    \caption{\textbf{Velocity deviations induced by an embedded planet, from analytical modelling of the wake.} Top: synthetic line-of-sight velocity maps for $^{12}$CO $J$=2-1 transition. Bottom: the corresponding residual velocities post-subtraction of a non-perturbed Keplerian power-law disc model, for both clockwise and counter-clockwise rotation of the disc. We model velocity deviations for embedded planetary masses of 1\,M$_{\mathrm{J}}$ and 10\,M$_{\mathrm{J}}$. All cubes are spatially and spectrally convolved to 0\farcs1 and 167\,ms$^{-1}$ respectively, to match the observations. The dotted rings represent the G1[38au] gap and the B1[26au] and B2[59au] dust rings.}
    \label{fig:wake_maps}
\end{figure*}

\subsection{Planetary mass estimate}
\label{sec:mp_est}

One possibility for velocity perturbations on the order of a few 10's of ms$^{-1}$ magnitude may be attributed to embedded planet(s). Given that an embedded planet would carve a gap via the exchange of angular momentum with the neighbouring fluid as a result of the tidal torque induced onto the planet by the wake. If the counter-balancing viscous torque is much lower than the rate of angular momentum exchange between the planet and neighbouring fluid, then a prominent gap-like structure is created. Therefore, the gap width and depth from observations can be directly used to infer the mass of the planetary body responsible for carving the gap \citep{Kanagawa2015, Dong2017}. The relation between gap depth, planet to star mass ratio, and disc properties is given by \citep{Kanagawa2015},
\begin{equation}
    \label{eq:Mp}
    \frac{\Sigma_{0}}{\Sigma_{\mathrm{gap}}} - 1 = 0.043 \left( \frac{M_{\mathrm{p}}}{M_{*}} \right)^{2} \left( \frac{h}{r} \right)^{-5} \alpha^{-1},
\end{equation}
where, $\Sigma_{\mathrm{gap}}$ and $\Sigma_0$ are the gas surface densities with and without depletion due to an embedded planet. $M_{\mathrm{p}}$ and $M_{*}$ are the planet and stellar masses, respectively. $\alpha$ is the disc viscosity. $h/r$ is the disc aspect ratio where the hydrostatic scale height ($h$) is given by Equation \ref{eq:h}.

Using $^{13}$CO as a temperature proxy within the G1 gap (at r$\approx$38au), a hydrostatic scale height of $\sim$2.95\,au is found for a temperature of 58K and stellar mass of 1.48\msun, giving a h/r of $\sim$0.08. From panel a) in Figure \ref{fig:pvp}, $\Sigma_{0}$ and $\Sigma_{\mathrm{gap}}$ are estimated to be $\sim$0.75\,g\,cm$^{-2}$ and 0.1\,g\,cm$^{-2}$ at r$\approx$38au, respectively. These measurements give an estimated planet mass of $\sim$1\,M$_{\mathrm{J}}$, for a disc viscosity ($\alpha$) of 10$^{-3}$ \citep[e.g.][]{Mulders2012, Ansdell2018}. This estimate of the planetary mass using the gas profile is comparable to the estimates made using dust emission in \cite{Dong2017} for the same disc viscosity and aspect ratio. Equivalently, this also translates to a thermal mass \citep{Goodman2001},
\begin{equation}
   m_{\mathrm{th}} = \frac{2}{3} \left( \frac{h_{\mathrm{p}}}{r_{\mathrm{p}}} \right)^{3} M_{*},
\end{equation}
of approximately 0.48M$_{\mathrm{J}}$, equivalent to half the planet mass. Estimated planetary masses are sensitive to variation in the thermal structure of the disc (refer to Equation \ref{eq:Mp}). Hence the analytics instead use thermal mass to estimate expected velocity perturbations without requiring a good constraint on the thermal structure.

\subsection{Comparison of the super-Keplerian arc to semi-analytical wake models}
\label{sec:wake}

In Section \ref{sec:mp_est} we deduced a planet mass of 1\,M$_{\mathrm{J}}$. Comparison plot of wake models in Figure \ref{fig:wake_maps} demonstrates that a co-planar 1\,M$_{\mathrm{J}}$ embedded body does not reproduce a signature large enough to explain the $\sim$75\,ms$^{-1}$ velocity deviations seen in the observations. Instead, from free exploration of the parameter space, we find kinematic deviations from a 10\,M$_{\mathrm{J}}$ mass planet are more consistent with the observations. This qualitative discrepancy in planet mass is likely the result of the near pole-on orientation of the disc. At an inclination of 13$^{\circ}$, the projected velocities in the observations may be dominated by the vertical velocity component, whilst the semi-analytic models only consider deviations in the plane of the disc (v$_{\mathrm{z}}$ = 0), thus over-predicting planetary masses. In this scenario, 10\,M$_{\mathrm{J}}$ serves as an upper limit rather than a proxy for the true planet mass.  

\cite{Ligi2018} and \cite{Gratton2019} have suggested clockwise rotation for the disc to be more consistent with the observed location of the point-like sources across time as detected in high-contrast imaging. This is in agreement with our observations of the $^{12}$CO velocity fields (Fig.\ref{fig:gv0+eddy_map_12co}), where the high velocity contours are found to slightly bend towards the West side of the minor axis. In comparison, we find the outer wake for a planet in prograde rotation to the disc (clockwise) extends in the opposite direction to the $^{12}$CO arc, and instead more closely resembles a Doppler-flip \citep{Casassus2019} rather than an extended arc-like structure. Whilst unlikely, we also consider retrograde motion. In this scenario, the orientation of the outer wake is found to be more in line with the observations, however there are still inconsistencies with the shape even for a 10\,M$_{\mathrm{J}}$.

%--------------------------------------------------
\section{Discussion}

\begin{figure*}
    \centering
    \includegraphics[width=0.9\textwidth]{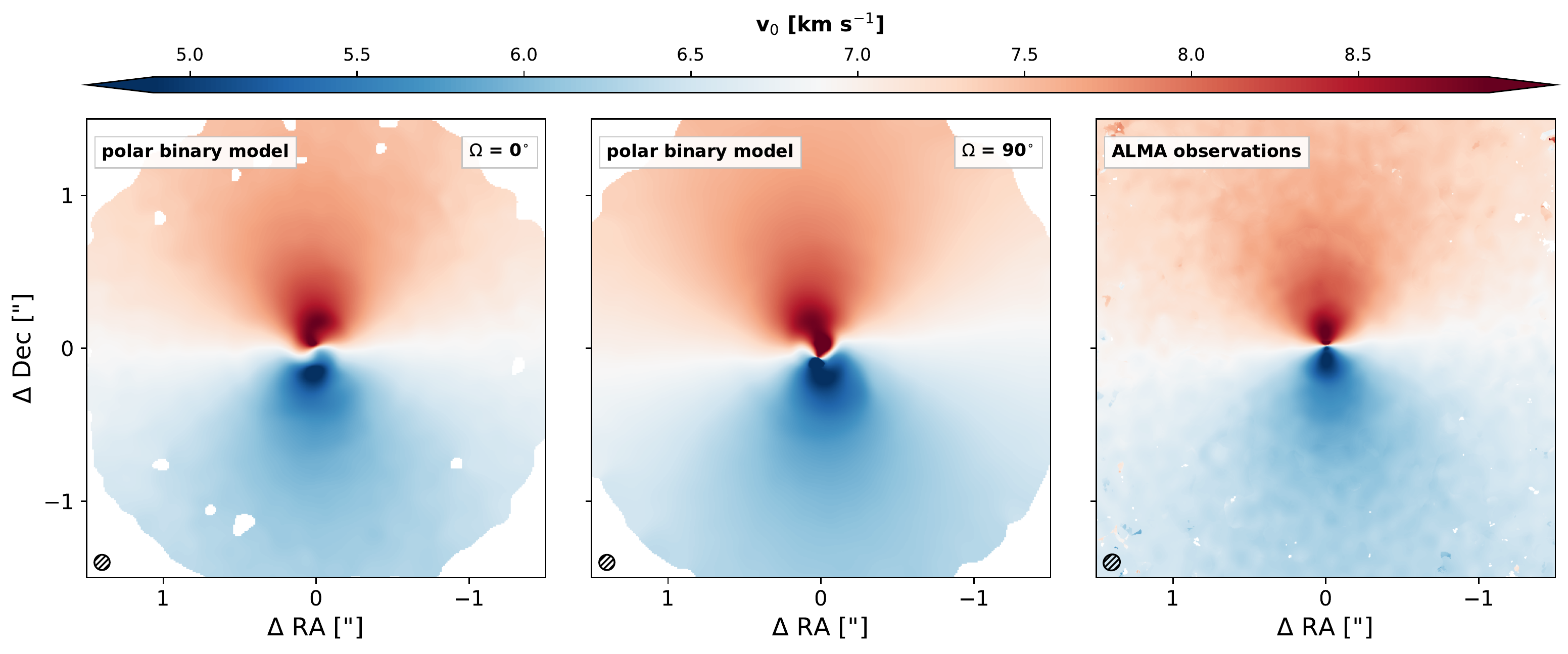}
    \caption{\textbf{A warp in the velocity fields of a polar binary model.} Left and middle panels: synthetic velocity fields maps of the polar binary model presented in \citet{Poblete2022} with an $\Omega$ of 0$^{\circ}$ and 90$^{\circ}$, respectively. Right panel: velocity fields map of the $^{12}$CO line emission shown in Fig. \ref{fig:gv0+eddy_map_12co}, top left panel.}
    \label{fig:hydro_comparison}
\end{figure*}

\subsection{Potential origins for the kinematic residual.}
\label{sec:discussion_planetmass}

\subsubsection{Is there an embedded planet in the gap?}
The relative depletion in the gas surface density profile infers a planetary mass of 1\,M$_{\mathrm{J}}$, consistent with planet mass estimates derived using dust emission \citep{Dong2017,Bae2018, Lodato2019}. However, a few caveats that still remain are:
\begin{itemize}
    \item uncertainty on the gas surface density profile. Ambiguity on the level of self-absorption between dust and gas emission in the disc results in an order of magnitude uncertainty on the gas surface density profile computed with and without continuum subtraction. Additionally, the use of molecular line emission relies on a scaling ratio between the molecular species used and H$_{2}$, which here has been pre-defined to equal ISM estimates. In planet forming discs several processes (isotope-selective photodissociation, freeze-out, and fractionation reactions) have been shown to drive relative isotopic ratios away from ISM estimates \citep{Miotello2014}. Collectively, these factors result in an order of magnitude uncertainty in the derived planetary mass from Equation \ref{eq:Mp} for a fixed disc aspect ratio and viscosity.
    \item uncertainty on the disc thermal structure and viscosity. Planet mass estimates from Equation \ref{eq:Mp} are sensitive to variations in the disc aspect ratio. To derive a planet mass of 10\,M$_{\mathrm{J}}$ using the current gas surface density profile, requires either an increase in the disc aspect ratio by a factor of $\sim$2.5 (h/r $\sim$0.2) or a disc viscosity of 10$^{-1}$. We can safely rule out a disc aspect ratio of 0.2 as a potential scenario, as this would in turn require a brightness temperature of $\sim$400\,K at an orbital radius of $\sim$38\,au (refer to Eq.\ref{eq:h}). For disc viscosities between 10$^{-3}$-10$^{-2}$ would result in a planet mass range of $\sim$1-3\,M$_{\mathrm{J}}$; comparable to the $\sim$1-4\,M$_{\mathrm{J}}$ estimate made for blob D in \cite{Gratton2019}.
    \item limited spatial resolution. The level of detail in observational data is intrinsically limited by the spatial resolution, spectral resolution and noise. The quality of the data also serves as a caveat in interpreting kinematic signatures. The effects of beam smearing have been suggested to produce artificial kinematic signatures \citep{Keppler2019, Boehler2021}. This effect is noted to be most prominent across regions of a map with steep brightness gradients. The arc in $^{12}$CO spans a region of the map that does not exhibit steep azimuthal variation in brightness. Thus ruling out any significant contamination by beam smearing in this scenario. However, to fully characterise the kinematic signature in $^{12}$CO, we suggest re-observing HD\,169142 at a higher spatial resolution (<\,1/3 the gap) and longer integration times (to achieve an SNR\,>\,8 at the location of the super-Keplerian arc).
\end{itemize}

\subsubsection{Misaligned discs?}
Near-IR polarised intensity images of HD\,169142 show signatures of shadowing across the outer composite dust rings along the major axis \citep{Quanz2013, Pohl2017, Bertrang2018, Rich2022}. One explanation for such shadows is subdivision and misalignment of the disc. Theoretical work by \cite{Young2022} demonstrate that subtraction of Keplerian models with a constant PA and inclination used throughout the disc can give rise to spurious spiral arms and/or central warps. In particular, their study produces multiple large arc-like features extending across a wide PA range, reminiscent of the kinematic excess observed in $^{12}$CO. Given the disagreement between the wake models to the observations, we propose disc misalignment as a plausible origin to the velocity residuals in Figure \ref{fig:gv0+eddy_map_12co}, however a full parameter space analysis remains beyond the scope of this paper.

\subsection{Is there a binary in the central cavity?}
\label{sec:circumbinary?}

Several propositions have been made as to the origin of the central cavity first seen in dust emission in \cite{Honda2012, Osorio2014, Fedele2017}. Theoretical work by \cite{Toci2020}, demonstrated that a few Jupiter mass planet placed close to the inner edge of the B1 dust ring could replicate a central cavity of a comparable size to the observations. Independently, \cite{Poblete2022} demonstrated that a stellar binary with a companion mass ratio of 0.1, semi-major axis 9.9\,au and eccentricity of 0.2, oriented at a 90$^{\circ}$ inclination (polar) to the plane of the disc, could reproduce not only a comparable cavity size but also the azimuthally uneven distribution of dust emission at mm-wavelengths for the B1 ring.

As shown in Section \ref{sec:COrings}, a central gas depleted cavity is present in all three gas tracers, including the optically thick $^{12}$CO $J$=2-1 emission. Whilst small scale velocity perturbations are observed between $\sim$50-75\,ms$^{-1}$ in magnitude, the line of sight velocity maps (refer to Fig. \ref{fig:gv0+eddy_map_12co}), show no large scale perturbations. Globally, the disc is found to be in near Keplerian orbit.

We compare the line of sight velocity fields from the proposed polar binary hydrodynamical model in \cite{Poblete2022}, to our observations. Two instances of the polar binary model are used, one with a longitude of the ascending node ($\Omega$) of 0$^{\circ}$ and another with 90$^{\circ}$. Both set of models are processed through {\sc mcfost} and convolved to a spatial and spectral resolutions of 0\farcs1 and 167\,ms$^{-1}$ respectively, to match the observations. The {\sc mcfost} products are then processed through {\sc bettermoments} to produce line of sight velocity maps using the {\sf Gaussian} method.

Velocity fields from the current polar binary configuration shows a central warp in the kinematics (Fig.\ref{fig:hydro_comparison}; in conflict with the observations. This suggests that either (i) the stellar mass companion ratio (0.1\,$\times$\,M$_{\mathrm{primary}}$) is too large; (ii) the current binary orientation induces greater velocity perturbations in the vertical direction than necessary; or (iii) the gas-density depletion in the central cavity is not due to a stellar binary.

%--------------------------------------------------
\section{Conclusion}

Our analysis of the ALMA band 6 observations of the disc around HD\,169142:
\begin{itemize}
    \item shows a central gas depleted cavity (R$_{\mathrm{cav}}\sim$22au) and annular gap [$\sim$38\,au]. This is accompanied by two concentric gas rings seen in all three ($^{12}$CO, $^{13}$CO and C$^{18}$O $J$=2-1) tracers (Fig.\ref{fig:COmaps,intensity}). The brightness temperature profile of all three tracers (including the optically thin tracer, C$^{18}$O) are found to remain above the CO freeze-out temperature of $\sim$20\,K within the inner 200\,au. Even at low disc altitudes, temperatures above sublimation favour a dynamical origin to the morphology of the gas structure, rather than a chemical one (Fig.\ref{fig:COprofiles}).
    \item finds a super-Keplerian kinematic excess azimuthally spanning across a PA range of -60 to 45$^{\circ}$ and radially in between the B1[26au] and B2[59au] dust rings (Fig.\ref{fig:gv0+eddy_map_12co}), resembling an arc-like structure. This velocity deviation is found to be in excess of background perturbations from localised pressure gradients and self-gravity (Fig.\ref{fig:vpr}. Part of this feature is found to closely overlap with the location of a high-intensity blob reported in \cite{Gratton2019}. Analytical models of planet-disk interaction do not appear to reproduce the observed kinematics, suggesting that if they are caused by a planet, the planet must be on an inclined orbit.
    \item demonstrates that the relative depletion in the gas surface density profile corresponds to a 1\,M$_{\mathrm{J}}$ planet (Fig.\ref{fig:col_den}; Section \ref{sec:mp_est}), consistent with the estimates made using dust emission \citep{Dong2017,Bae2018, Lodato2019}. 
    \item shows relatively smooth kinematics within the central cavity. In comparison, the current polar binary model proposed in \cite{Poblete2022} results in resolved warping of the high velocity fields (Fig.\ref{fig:hydro_comparison}). Therefore, we reconsider this model as being responsible for depleting the gas and dust content in the central cavity. 
\end{itemize}

\section*{Acknowledgements}
H. Garg, C. Pinte, I. Hammond and D.~J. Price acknowledge funding from the Australian Research Council via grant DP180104235. C. Pinte also acknowledges funding from grant FT170100040. J. Calcino acknowledged support the LANL LDRD program. Approved for released as xxxx. V. Christiaens acknowledges support from the Belgian FRS-FNRS. This paper makes use of the following ALMA data: ADS/JAO.ALMA\#2015.1.00490.S and ADS/JAO.ALMA\#2016.1.00344.S. ALMA is a partnership of ESO (representing its member states), NSF (USA), and NINS (Japan), together with NRC (Canada), MOST and ASIAA (Taiwan), and KASI (Republic of Korea), in cooperation with the Republic of Chile. The Joint ALMA Observatory is operated by ESO, AUI/NRAO, and NAOJ. The National Radio Astronomy Observatory is a facility of the National Science Foundation operated under cooperative agreement by Associated Universities, Inc. This paper makes use of the following publicly available programs: {\sc mcfost}, \url{https://github.com/cpinte/mcfost}; {\sc pymcfost}, \url{https://github.com/cpinte/pymcfost}; {\sc bettermoments}, \url{https://github.com/richteague/bettermoments}; {\sc eddy}, \url{https://github.com/richteague/eddy}; and {\sc wakeflow}, \url{https://github.com/TomHilder/wakeflow}. The polar binary {\sc phantom} models were provided by Pedro~P. Poblete. Finally, we thank Sebasti\'{a}n P\'{e}rez for helpful discussions.

%%%%%%%%%%%%%%%%%%%%%%%%%%%%%%%%%%%%%%%%%%%%%%%%%%
\section*{Data Availability}
The original raw data sets are publicly available on the ALMA archive under Project codes 2015.1.00490.S and 2016.1.00344.S. The self-calibrated products are available via request from the corresponding author.
%The inclusion of a Data Availability Statement is a requirement for articles published in MNRAS. Data Availability Statements provide a standardised format for readers to understand the availability of data underlying the research results described in the article. The statement may refer to original data generated in the course of the study or to third-party data analysed in the article. The statement should describe and provide means of access, where possible, by linking to the data or providing the required accession numbers for the relevant databases or DOIs.

%%%%%%%%%%%%%%%%%%%% REFERENCES %%%%%%%%%%%%%%%%%%

% The best way to enter references is to use BibTeX:

\bibliographystyle{mnras}
\bibliography{refs} % if your bibtex file is called example.bib

\begin{thebibliography}{}
\makeatletter
\relax
\def\mn@urlcharsother{\let\do\@makeother \do\$\do\&\do\#\do\^\do\_\do\%\do\~}
\def\mn@doi{\begingroup\mn@urlcharsother \@ifnextchar [ {\mn@doi@}
  {\mn@doi@[]}}
\def\mn@doi@[#1]#2{\def\@tempa{#1}\ifx\@tempa\@empty \href
  {http://dx.doi.org/#2} {doi:#2}\else \href {http://dx.doi.org/#2} {#1}\fi
  \endgroup}
\def\mn@eprint#1#2{\mn@eprint@#1:#2::\@nil}
\def\mn@eprint@arXiv#1{\href {http://arxiv.org/abs/#1} {{\tt arXiv:#1}}}
\def\mn@eprint@dblp#1{\href {http://dblp.uni-trier.de/rec/bibtex/#1.xml}
  {dblp:#1}}
\def\mn@eprint@#1:#2:#3:#4\@nil{\def\@tempa {#1}\def\@tempb {#2}\def\@tempc
  {#3}\ifx \@tempc \@empty \let \@tempc \@tempb \let \@tempb \@tempa \fi \ifx
  \@tempb \@empty \def\@tempb {arXiv}\fi \@ifundefined
  {mn@eprint@\@tempb}{\@tempb:\@tempc}{\expandafter \expandafter \csname
  mn@eprint@\@tempb\endcsname \expandafter{\@tempc}}}

\bibitem[\protect\citeauthoryear{{ALMA Partnership} et~al.,}{{ALMA Partnership}
  et~al.}{2015}]{ALMApartnership}
{ALMA Partnership} et~al., 2015, \mn@doi [\apjl] {10.1088/2041-8205/808/1/L3},
  \href {https://ui.adsabs.harvard.edu/abs/2015ApJ...808L...3A} {808, L3}

\bibitem[\protect\citeauthoryear{{Andrews} et~al.,}{{Andrews}
  et~al.}{2016}]{Andrews2016}
{Andrews} S.~M.,  et~al., 2016, \mn@doi [\apjl] {10.3847/2041-8205/820/2/L40},
  \href {https://ui.adsabs.harvard.edu/abs/2016ApJ...820L..40A} {820, L40}

\bibitem[\protect\citeauthoryear{{Andrews} et~al.,}{{Andrews}
  et~al.}{2018}]{Andrews2018}
{Andrews} S.~M.,  et~al., 2018, \mn@doi [\apjl] {10.3847/2041-8213/aaf741},
  \href {https://ui.adsabs.harvard.edu/abs/2018ApJ...869L..41A} {869, L41}

\bibitem[\protect\citeauthoryear{{Ansdell} et~al.,}{{Ansdell}
  et~al.}{2018}]{Ansdell2018}
{Ansdell} M.,  et~al., 2018, \mn@doi [\apj] {10.3847/1538-4357/aab890}, \href
  {https://ui.adsabs.harvard.edu/abs/2018ApJ...859...21A} {859, 21}

\bibitem[\protect\citeauthoryear{{Ayliffe}, {Laibe}, {Price}  \&
  {Bate}}{{Ayliffe} et~al.}{2012}]{Ayliffe2012}
{Ayliffe} B.~A.,  {Laibe} G.,  {Price} D.~J.,   {Bate} M.~R.,  2012, \mn@doi
  [\mnras] {10.1111/j.1365-2966.2012.20967.x}, \href
  {https://ui.adsabs.harvard.edu/abs/2012MNRAS.423.1450A} {423, 1450}

\bibitem[\protect\citeauthoryear{{Bae}, {Pinilla}  \& {Birnstiel}}{{Bae}
  et~al.}{2018}]{Bae2018}
{Bae} J.,  {Pinilla} P.,   {Birnstiel} T.,  2018, \mn@doi [\apjl]
  {10.3847/2041-8213/aadd51}, \href
  {https://ui.adsabs.harvard.edu/abs/2018ApJ...864L..26B} {864, L26}

\bibitem[\protect\citeauthoryear{{Bertrang}, {Avenhaus}, {Casassus},
  {Montesinos}, {Kirchschlager}, {Perez}, {Cieza}  \& {Wolf}}{{Bertrang}
  et~al.}{2018}]{Bertrang2018}
{Bertrang} G.~H.~M.,  {Avenhaus} H.,  {Casassus} S.,  {Montesinos} M.,
  {Kirchschlager} F.,  {Perez} S.,  {Cieza} L.,   {Wolf} S.,  2018, \mn@doi
  [\mnras] {10.1093/mnras/stx3052}, \href
  {https://ui.adsabs.harvard.edu/abs/2018MNRAS.474.5105B} {474, 5105}

\bibitem[\protect\citeauthoryear{{Biller} et~al.,}{{Biller}
  et~al.}{2014}]{Biller2014}
{Biller} B.~A.,  et~al., 2014, \mn@doi [\apjl] {10.1088/2041-8205/792/1/L22},
  \href {https://ui.adsabs.harvard.edu/abs/2014ApJ...792L..22B} {792, L22}

\bibitem[\protect\citeauthoryear{{Blondel} \& {Djie}}{{Blondel} \&
  {Djie}}{2006}]{Blondel2006}
{Blondel} P.~F.~C.,  {Djie} H.~R.~E. T.~A.,  2006, \mn@doi [\aap]
  {10.1051/0004-6361:20040269}, \href
  {https://ui.adsabs.harvard.edu/abs/2006A&A...456.1045B} {456, 1045}

\bibitem[\protect\citeauthoryear{{Boehler} et~al.,}{{Boehler}
  et~al.}{2021}]{Boehler2021}
{Boehler} Y.,  et~al., 2021, \mn@doi [\aap] {10.1051/0004-6361/202040089},
  \href {https://ui.adsabs.harvard.edu/abs/2021A&A...650A..59B} {650, A59}

\bibitem[\protect\citeauthoryear{{Bollati}, {Lodato}, {Price}  \&
  {Pinte}}{{Bollati} et~al.}{2021}]{Bollati2021}
{Bollati} F.,  {Lodato} G.,  {Price} D.~J.,   {Pinte} C.,  2021, \mn@doi
  [\mnras] {10.1093/mnras/stab1145}, \href
  {https://ui.adsabs.harvard.edu/abs/2021MNRAS.504.5444B} {504, 5444}

\bibitem[\protect\citeauthoryear{{Casassus} \& {P{\'e}rez}}{{Casassus} \&
  {P{\'e}rez}}{2019}]{Casassus2019}
{Casassus} S.,  {P{\'e}rez} S.,  2019, \mn@doi [\apjl]
  {10.3847/2041-8213/ab4425}, \href
  {https://ui.adsabs.harvard.edu/abs/2019ApJ...883L..41C} {883, L41}

\bibitem[\protect\citeauthoryear{{Casassus} et~al.,}{{Casassus}
  et~al.}{2021}]{Casassus2021}
{Casassus} S.,  et~al., 2021, \mn@doi [\mnras] {10.1093/mnras/stab2359}, \href
  {https://ui.adsabs.harvard.edu/abs/2021MNRAS.tmp.2140C} {}

\bibitem[\protect\citeauthoryear{{Dipierro}, {Price}, {Laibe}, {Hirsh},
  {Cerioli}  \& {Lodato}}{{Dipierro} et~al.}{2015}]{Dipierro2015}
{Dipierro} G.,  {Price} D.,  {Laibe} G.,  {Hirsh} K.,  {Cerioli} A.,   {Lodato}
  G.,  2015, \mn@doi [\mnras] {10.1093/mnrasl/slv105}, \href
  {https://ui.adsabs.harvard.edu/abs/2015MNRAS.453L..73D} {453, L73}

\bibitem[\protect\citeauthoryear{{Dong} \& {Fung}}{{Dong} \&
  {Fung}}{2017}]{Dong2017}
{Dong} R.,  {Fung} J.,  2017, \mn@doi [\apj] {10.3847/1538-4357/835/2/146},
  \href {https://ui.adsabs.harvard.edu/abs/2017ApJ...835..146D} {835, 146}

\bibitem[\protect\citeauthoryear{{Dong}, {Fung}  \& {Chiang}}{{Dong}
  et~al.}{2016}]{Dong2016}
{Dong} R.,  {Fung} J.,   {Chiang} E.,  2016, \mn@doi [\apj]
  {10.3847/0004-637X/826/1/75}, \href
  {https://ui.adsabs.harvard.edu/abs/2016ApJ...826...75D} {826, 75}

\bibitem[\protect\citeauthoryear{{Dunkin}, {Barlow}  \& {Ryan}}{{Dunkin}
  et~al.}{1997}]{Dunkin1997}
{Dunkin} S.~K.,  {Barlow} M.~J.,   {Ryan} S.~G.,  1997, \mn@doi [\mnras]
  {10.1093/mnras/286.3.604}, \href
  {https://ui.adsabs.harvard.edu/abs/1997MNRAS.286..604D} {286, 604}

\bibitem[\protect\citeauthoryear{{Fedele} et~al.,}{{Fedele}
  et~al.}{2017}]{Fedele2017}
{Fedele} D.,  et~al., 2017, \mn@doi [\aap] {10.1051/0004-6361/201629860}, \href
  {https://ui.adsabs.harvard.edu/abs/2017A&A...600A..72F} {600, A72}

\bibitem[\protect\citeauthoryear{{Flock}, {Ruge}, {Dzyurkevich}, {Henning},
  {Klahr}  \& {Wolf}}{{Flock} et~al.}{2015}]{Flock2015}
{Flock} M.,  {Ruge} J.~P.,  {Dzyurkevich} N.,  {Henning} T.,  {Klahr} H.,
  {Wolf} S.,  2015, \mn@doi [\aap] {10.1051/0004-6361/201424693}, \href
  {https://ui.adsabs.harvard.edu/abs/2015A&A...574A..68F} {574, A68}

\bibitem[\protect\citeauthoryear{{Gaia Collaboration} et~al.,}{{Gaia
  Collaboration} et~al.}{2016}]{Gaia2016}
{Gaia Collaboration} et~al., 2016, \mn@doi [\aap]
  {10.1051/0004-6361/201629512}, \href
  {https://ui.adsabs.harvard.edu/abs/2016A&A...595A...2G} {595, A2}

\bibitem[\protect\citeauthoryear{{Garg} et~al.,}{{Garg}
  et~al.}{2021}]{Garg2021}
{Garg} H.,  et~al., 2021, \mn@doi [\mnras] {10.1093/mnras/stab800}, \href
  {https://ui.adsabs.harvard.edu/abs/2021MNRAS.504..782G} {504, 782}

\bibitem[\protect\citeauthoryear{{Goodman} \& {Rafikov}}{{Goodman} \&
  {Rafikov}}{2001}]{Goodman2001}
{Goodman} J.,  {Rafikov} R.~R.,  2001, \mn@doi [\apj] {10.1086/320572}, \href
  {https://ui.adsabs.harvard.edu/abs/2001ApJ...552..793G} {552, 793}

\bibitem[\protect\citeauthoryear{{Gratton} et~al.,}{{Gratton}
  et~al.}{2019}]{Gratton2019}
{Gratton} R.,  et~al., 2019, \mn@doi [\aap] {10.1051/0004-6361/201834760},
  \href {https://ui.adsabs.harvard.edu/abs/2019A&A...623A.140G} {623, A140}

\bibitem[\protect\citeauthoryear{{Honda} et~al.,}{{Honda}
  et~al.}{2012}]{Honda2012}
{Honda} M.,  et~al., 2012, \mn@doi [\apj] {10.1088/0004-637X/752/2/143}, \href
  {https://ui.adsabs.harvard.edu/abs/2012ApJ...752..143H} {752, 143}

\bibitem[\protect\citeauthoryear{{Kanagawa}, {Muto}, {Tanaka}, {Tanigawa},
  {Takeuchi}, {Tsukagoshi}  \& {Momose}}{{Kanagawa}
  et~al.}{2015}]{Kanagawa2015}
{Kanagawa} K.~D.,  {Muto} T.,  {Tanaka} H.,  {Tanigawa} T.,  {Takeuchi} T.,
  {Tsukagoshi} T.,   {Momose} M.,  2015, \mn@doi [\apjl]
  {10.1088/2041-8205/806/1/L15}, \href
  {https://ui.adsabs.harvard.edu/abs/2015ApJ...806L..15K} {806, L15}

\bibitem[\protect\citeauthoryear{{Kastner} et~al.,}{{Kastner}
  et~al.}{2018}]{Kastner2018}
{Kastner} J.~H.,  et~al., 2018, \mn@doi [\apj] {10.3847/1538-4357/aacff7},
  \href {https://ui.adsabs.harvard.edu/abs/2018ApJ...863..106K} {863, 106}

\bibitem[\protect\citeauthoryear{{Keppler} et~al.,}{{Keppler}
  et~al.}{2019}]{Keppler2019}
{Keppler} M.,  et~al., 2019, \mn@doi [\aap] {10.1051/0004-6361/201935034},
  \href {https://ui.adsabs.harvard.edu/abs/2019A&A...625A.118K} {625, A118}

\bibitem[\protect\citeauthoryear{{Kretke} \& {Lin}}{{Kretke} \&
  {Lin}}{2007}]{Kretke2007}
{Kretke} K.~A.,  {Lin} D.~N.~C.,  2007, \mn@doi [\apjl] {10.1086/520718}, \href
  {https://ui.adsabs.harvard.edu/abs/2007ApJ...664L..55K} {664, L55}

\bibitem[\protect\citeauthoryear{{Ligi} et~al.,}{{Ligi}
  et~al.}{2018}]{Ligi2018}
{Ligi} R.,  et~al., 2018, \mn@doi [\mnras] {10.1093/mnras/stx2318}, \href
  {https://ui.adsabs.harvard.edu/abs/2018MNRAS.473.1774L} {473, 1774}

\bibitem[\protect\citeauthoryear{{Lodato} et~al.,}{{Lodato}
  et~al.}{2019}]{Lodato2019}
{Lodato} G.,  et~al., 2019, \mn@doi [\mnras] {10.1093/mnras/stz913}, \href
  {https://ui.adsabs.harvard.edu/abs/2019MNRAS.486..453L} {486, 453}

\bibitem[\protect\citeauthoryear{{Lyo}, {Ohashi}, {Qi}, {Wilner}  \&
  {Su}}{{Lyo} et~al.}{2011}]{Lyo2011}
{Lyo} A.~R.,  {Ohashi} N.,  {Qi} C.,  {Wilner} D.~J.,   {Su} Y.-N.,  2011,
  \mn@doi [\aj] {10.1088/0004-6256/142/5/151}, \href
  {https://ui.adsabs.harvard.edu/abs/2011AJ....142..151L} {142, 151}

\bibitem[\protect\citeauthoryear{{Mac{\'\i}as} et~al.,}{{Mac{\'\i}as}
  et~al.}{2019}]{Macias2019}
{Mac{\'\i}as} E.,  et~al., 2019, \mn@doi [\apj] {10.3847/1538-4357/ab31a2},
  \href {https://ui.adsabs.harvard.edu/abs/2019ApJ...881..159M} {881, 159}

\bibitem[\protect\citeauthoryear{{Mangum} \& {Shirley}}{{Mangum} \&
  {Shirley}}{2015}]{Mangum2015}
{Mangum} J.~G.,  {Shirley} Y.~L.,  2015, \mn@doi [\pasp] {10.1086/680323},
  \href {https://ui.adsabs.harvard.edu/abs/2015PASP..127..266M} {127, 266}

\bibitem[\protect\citeauthoryear{{Miotello}, {Bruderer}  \& {van
  Dishoeck}}{{Miotello} et~al.}{2014}]{Miotello2014}
{Miotello} A.,  {Bruderer} S.,   {van Dishoeck} E.~F.,  2014, \mn@doi [\aap]
  {10.1051/0004-6361/201424712}, \href
  {https://ui.adsabs.harvard.edu/abs/2014A&A...572A..96M} {572, A96}

\bibitem[\protect\citeauthoryear{{Momose} et~al.,}{{Momose}
  et~al.}{2015}]{Momose2015}
{Momose} M.,  et~al., 2015, \mn@doi [\pasj] {10.1093/pasj/psv051}, \href
  {https://ui.adsabs.harvard.edu/abs/2015PASJ...67...83M} {67, 83}

\bibitem[\protect\citeauthoryear{{Mulders} \& {Dominik}}{{Mulders} \&
  {Dominik}}{2012}]{Mulders2012}
{Mulders} G.~D.,  {Dominik} C.,  2012, \mn@doi [\aap]
  {10.1051/0004-6361/201118127}, \href
  {https://ui.adsabs.harvard.edu/abs/2012A&A...539A...9M} {539, A9}

\bibitem[\protect\citeauthoryear{{Okuzumi}, {Momose}, {Sirono}, {Kobayashi}  \&
  {Tanaka}}{{Okuzumi} et~al.}{2016}]{Okuzumi2016}
{Okuzumi} S.,  {Momose} M.,  {Sirono} S.-i.,  {Kobayashi} H.,   {Tanaka} H.,
  2016, \mn@doi [\apj] {10.3847/0004-637X/821/2/82}, \href
  {https://ui.adsabs.harvard.edu/abs/2016ApJ...821...82O} {821, 82}

\bibitem[\protect\citeauthoryear{{Osorio} et~al.,}{{Osorio}
  et~al.}{2014}]{Osorio2014}
{Osorio} M.,  et~al., 2014, \mn@doi [\apjl] {10.1088/2041-8205/791/2/L36},
  \href {https://ui.adsabs.harvard.edu/abs/2014ApJ...791L..36O} {791, L36}

\bibitem[\protect\citeauthoryear{{Pani{\'c}}, {Hogerheijde}, {Wilner}  \&
  {Qi}}{{Pani{\'c}} et~al.}{2008}]{Panic2008}
{Pani{\'c}} O.,  {Hogerheijde} M.~R.,  {Wilner} D.,   {Qi} C.,  2008, \mn@doi
  [\aap] {10.1051/0004-6361:20079261}, \href
  {https://ui.adsabs.harvard.edu/abs/2008A&A...491..219P} {491, 219}

\bibitem[\protect\citeauthoryear{{Perez} et~al.,}{{Perez}
  et~al.}{2015}]{Perez2015}
{Perez} S.,  et~al., 2015, \mn@doi [\apj] {10.1088/0004-637X/798/2/85}, \href
  {https://ui.adsabs.harvard.edu/abs/2015ApJ...798...85P} {798, 85}

\bibitem[\protect\citeauthoryear{{P{\'e}rez}, {Casassus}  \&
  {Ben{\'\i}tez-Llambay}}{{P{\'e}rez} et~al.}{2018}]{Perez2018}
{P{\'e}rez} S.,  {Casassus} S.,   {Ben{\'\i}tez-Llambay} P.,  2018, \mn@doi
  [\mnras] {10.1093/mnrasl/sly109}, \href
  {https://ui.adsabs.harvard.edu/abs/2018MNRAS.480L..12P} {480, L12}

\bibitem[\protect\citeauthoryear{{P{\'e}rez}, {Casassus}, {Baruteau}, {Dong},
  {Hales}  \& {Cieza}}{{P{\'e}rez} et~al.}{2019}]{Perez2019}
{P{\'e}rez} S.,  {Casassus} S.,  {Baruteau} C.,  {Dong} R.,  {Hales} A.,
  {Cieza} L.,  2019, \mn@doi [\aj] {10.3847/1538-3881/ab1f88}, \href
  {https://ui.adsabs.harvard.edu/abs/2019AJ....158...15P} {158, 15}

\bibitem[\protect\citeauthoryear{{Pinilla}, {Benisty}  \&
  {Birnstiel}}{{Pinilla} et~al.}{2012}]{Pinilla2012}
{Pinilla} P.,  {Benisty} M.,   {Birnstiel} T.,  2012, \mn@doi [\aap]
  {10.1051/0004-6361/201219315}, \href
  {https://ui.adsabs.harvard.edu/abs/2012A&A...545A..81P} {545, A81}

\bibitem[\protect\citeauthoryear{{Pinilla}, {Flock}, {Ovelar}  \&
  {Birnstiel}}{{Pinilla} et~al.}{2016}]{Pinilla2016}
{Pinilla} P.,  {Flock} M.,  {Ovelar} M. d.~J.,   {Birnstiel} T.,  2016, \mn@doi
  [\aap] {10.1051/0004-6361/201628441}, \href
  {https://ui.adsabs.harvard.edu/abs/2016A&A...596A..81P} {596, A81}

\bibitem[\protect\citeauthoryear{{Pinte}, {M{\'e}nard}, {Duch{\^e}ne}  \&
  {Bastien}}{{Pinte} et~al.}{2006}]{Pinte2006mcfost}
{Pinte} C.,  {M{\'e}nard} F.,  {Duch{\^e}ne} G.,   {Bastien} P.,  2006, \mn@doi
  [\aap] {10.1051/0004-6361:20053275}, \href
  {https://ui.adsabs.harvard.edu/abs/2006A&A...459..797P} {459, 797}

\bibitem[\protect\citeauthoryear{{Pinte}, {Harries}, {Min}, {Watson},
  {Dullemond}, {Woitke}, {M{\'e}nard}  \& {Dur{\'a}n-Rojas}}{{Pinte}
  et~al.}{2009}]{Pinte2009mcfost}
{Pinte} C.,  {Harries} T.~J.,  {Min} M.,  {Watson} A.~M.,  {Dullemond} C.~P.,
  {Woitke} P.,  {M{\'e}nard} F.,   {Dur{\'a}n-Rojas} M.~C.,  2009, \mn@doi
  [\aap] {10.1051/0004-6361/200811555}, \href
  {https://ui.adsabs.harvard.edu/abs/2009A&A...498..967P} {498, 967}

\bibitem[\protect\citeauthoryear{{Pinte} et~al.,}{{Pinte}
  et~al.}{2018}]{Pinte2018}
{Pinte} C.,  et~al., 2018, \mn@doi [\aap] {10.1051/0004-6361/201731377}, \href
  {https://ui.adsabs.harvard.edu/abs/2018A&A...609A..47P} {609, A47}

\bibitem[\protect\citeauthoryear{{Pinte} et~al.,}{{Pinte}
  et~al.}{2019}]{Pinte2019}
{Pinte} C.,  et~al., 2019, \mn@doi [Nature Astronomy]
  {10.1038/s41550-019-0852-6}, \href
  {https://ui.adsabs.harvard.edu/abs/2019NatAs...3.1109P} {3, 1109}

\bibitem[\protect\citeauthoryear{{Poblete} et~al.,}{{Poblete}
  et~al.}{2022}]{Poblete2022}
{Poblete} P.~P.,  et~al., 2022, \mn@doi [\mnras] {10.1093/mnras/stab3474},
  \href {https://ui.adsabs.harvard.edu/abs/2022MNRAS.510..205P} {510, 205}

\bibitem[\protect\citeauthoryear{{Pohl} et~al.,}{{Pohl}
  et~al.}{2017}]{Pohl2017}
{Pohl} A.,  et~al., 2017, \mn@doi [\apj] {10.3847/1538-4357/aa94c2}, \href
  {https://ui.adsabs.harvard.edu/abs/2017ApJ...850...52P} {850, 52}

\bibitem[\protect\citeauthoryear{{Quanz}, {Avenhaus}, {Buenzli}, {Garufi},
  {Schmid}  \& {Wolf}}{{Quanz} et~al.}{2013}]{Quanz2013}
{Quanz} S.~P.,  {Avenhaus} H.,  {Buenzli} E.,  {Garufi} A.,  {Schmid} H.~M.,
  {Wolf} S.,  2013, \mn@doi [\apjl] {10.1088/2041-8205/766/1/L2}, \href
  {https://ui.adsabs.harvard.edu/abs/2013ApJ...766L...2Q} {766, L2}

\bibitem[\protect\citeauthoryear{{Rafikov}}{{Rafikov}}{2002}]{Rafikov2002}
{Rafikov} R.~R.,  2002, \mn@doi [\apj] {10.1086/339399}, \href
  {https://ui.adsabs.harvard.edu/abs/2002ApJ...569..997R} {569, 997}

\bibitem[\protect\citeauthoryear{{Raman}, {Lisanti}, {Wilner}, {Qi}  \&
  {Hogerheijde}}{{Raman} et~al.}{2006}]{Raman2006}
{Raman} A.,  {Lisanti} M.,  {Wilner} D.~J.,  {Qi} C.,   {Hogerheijde} M.,
  2006, \mn@doi [\aj] {10.1086/500587}, \href
  {https://ui.adsabs.harvard.edu/abs/2006AJ....131.2290R} {131, 2290}

\bibitem[\protect\citeauthoryear{{Reggiani} et~al.,}{{Reggiani}
  et~al.}{2014}]{Reggiani2014}
{Reggiani} M.,  et~al., 2014, \mn@doi [\apjl] {10.1088/2041-8205/792/1/L23},
  \href {https://ui.adsabs.harvard.edu/abs/2014ApJ...792L..23R} {792, L23}

\bibitem[\protect\citeauthoryear{{Rich} et~al.,}{{Rich}
  et~al.}{2022}]{Rich2022}
{Rich} E.~A.,  et~al., 2022, \mn@doi [\aj] {10.3847/1538-3881/ac7be4}, \href
  {https://ui.adsabs.harvard.edu/abs/2022AJ....164..109R} {164, 109}

\bibitem[\protect\citeauthoryear{{Rosenfeld}, {Andrews}, {Hughes}, {Wilner}  \&
  {Qi}}{{Rosenfeld} et~al.}{2013}]{Rosenfeld2013}
{Rosenfeld} K.~A.,  {Andrews} S.~M.,  {Hughes} A.~M.,  {Wilner} D.~J.,   {Qi}
  C.,  2013, \mn@doi [\apj] {10.1088/0004-637X/774/1/16}, \href
  {https://ui.adsabs.harvard.edu/abs/2013ApJ...774...16R} {774, 16}

\bibitem[\protect\citeauthoryear{{Rosotti}, {Juhasz}, {Booth}  \&
  {Clarke}}{{Rosotti} et~al.}{2016}]{Rosotti2016}
{Rosotti} G.~P.,  {Juhasz} A.,  {Booth} R.~A.,   {Clarke} C.~J.,  2016, \mn@doi
  [\mnras] {10.1093/mnras/stw691}, \href
  {https://ui.adsabs.harvard.edu/abs/2016MNRAS.459.2790R} {459, 2790}

\bibitem[\protect\citeauthoryear{{Saito} \& {Sirono}}{{Saito} \&
  {Sirono}}{2011}]{Saito2011}
{Saito} E.,  {Sirono} S.-i.,  2011, \mn@doi [\apj]
  {10.1088/0004-637X/728/1/20}, \href
  {https://ui.adsabs.harvard.edu/abs/2011ApJ...728...20S} {728, 20}

\bibitem[\protect\citeauthoryear{{Sch{\"o}ier}, {van der Tak}, {van Dishoeck}
  \& {Black}}{{Sch{\"o}ier} et~al.}{2005}]{Schoier2005}
{Sch{\"o}ier} F.~L.,  {van der Tak} F.~F.~S.,  {van Dishoeck} E.~F.,   {Black}
  J.~H.,  2005, \mn@doi [\aap] {10.1051/0004-6361:20041729}, \href
  {https://ui.adsabs.harvard.edu/abs/2005A&A...432..369S} {432, 369}

\bibitem[\protect\citeauthoryear{{Schwarz}, {Bergin}, {Cleeves}, {Blake},
  {Zhang}, {{\"O}berg}, {van Dishoeck}  \& {Qi}}{{Schwarz}
  et~al.}{2016}]{Schwarz2016}
{Schwarz} K.~R.,  {Bergin} E.~A.,  {Cleeves} L.~I.,  {Blake} G.~A.,  {Zhang}
  K.,  {{\"O}berg} K.~I.,  {van Dishoeck} E.~F.,   {Qi} C.,  2016, \mn@doi
  [\apj] {10.3847/0004-637X/823/2/91}, \href
  {https://ui.adsabs.harvard.edu/abs/2016ApJ...823...91S} {823, 91}

\bibitem[\protect\citeauthoryear{{Sierra} et~al.,}{{Sierra}
  et~al.}{2021}]{Sierra2021}
{Sierra} A.,  et~al., 2021, \mn@doi [\apjs] {10.3847/1538-4365/ac1431}, \href
  {https://ui.adsabs.harvard.edu/abs/2021ApJS..257...14S} {257, 14}

\bibitem[\protect\citeauthoryear{{Stahl}, {Casassus}  \& {Wilson}}{{Stahl}
  et~al.}{2008}]{Stahl2008}
{Stahl} O.,  {Casassus} S.,   {Wilson} T.,  2008, \mn@doi [\aap]
  {10.1051/0004-6361:20078747}, \href
  {https://ui.adsabs.harvard.edu/abs/2008A&A...477..865S} {477, 865}

\bibitem[\protect\citeauthoryear{Teague}{Teague}{2019}]{eddy}
Teague R.,  2019, \mn@doi [The Journal of Open Source Software]
  {10.21105/joss.01220}, 4, 1220

\bibitem[\protect\citeauthoryear{{Teague} \& {Foreman-Mackey}}{{Teague} \&
  {Foreman-Mackey}}{2018}]{bettermoments}
{Teague} R.,  {Foreman-Mackey} D.,  2018, \mn@doi [Research Notes of the
  American Astronomical Society] {10.3847/2515-5172/aae265}, \href
  {https://ui.adsabs.harvard.edu/abs/2018RNAAS...2c.173T} {2, 173}

\bibitem[\protect\citeauthoryear{{Teague}, {Bae}, {Bergin}, {Birnstiel}  \&
  {Foreman-Mackey}}{{Teague} et~al.}{2018}]{Teague2018}
{Teague} R.,  {Bae} J.,  {Bergin} E.~A.,  {Birnstiel} T.,   {Foreman-Mackey}
  D.,  2018, \mn@doi [\apjl] {10.3847/2041-8213/aac6d7}, \href
  {https://ui.adsabs.harvard.edu/abs/2018ApJ...860L..12T} {860, L12}

\bibitem[\protect\citeauthoryear{{Teague} et~al.,}{{Teague}
  et~al.}{2022}]{Teague2022}
{Teague} R.,  et~al., 2022, arXiv e-prints, \href
  {https://ui.adsabs.harvard.edu/abs/2022arXiv220804837T} {p. arXiv:2208.04837}

\bibitem[\protect\citeauthoryear{{Toci}, {Lodato}, {Fedele}, {Testi}  \&
  {Pinte}}{{Toci} et~al.}{2020}]{Toci2020}
{Toci} C.,  {Lodato} G.,  {Fedele} D.,  {Testi} L.,   {Pinte} C.,  2020,
  \mn@doi [\apjl] {10.3847/2041-8213/ab5c87}, \href
  {https://ui.adsabs.harvard.edu/abs/2020ApJ...888L...4T} {888, L4}

\bibitem[\protect\citeauthoryear{{Veronesi} et~al.,}{{Veronesi}
  et~al.}{2020}]{Veronesi2020}
{Veronesi} B.,  et~al., 2020, \mn@doi [\mnras] {10.1093/mnras/staa1278}, \href
  {https://ui.adsabs.harvard.edu/abs/2020MNRAS.495.1913V} {495, 1913}

\bibitem[\protect\citeauthoryear{{Weingartner} \& {Draine}}{{Weingartner} \&
  {Draine}}{2001}]{Weingartner01}
{Weingartner} J.~C.,  {Draine} B.~T.,  2001, \mn@doi [\apj] {10.1086/318651},
  \href
  {http://adsabs.harvard.edu/cgi-bin/nph-bib_query?bibcode=2001ApJ...548..296W&db_key=AST}
  {548, 296}

\bibitem[\protect\citeauthoryear{{Wilson} \& {Rood}}{{Wilson} \&
  {Rood}}{1994}]{Wilson1994}
{Wilson} T.~L.,  {Rood} R.,  1994, \mn@doi [\araa]
  {10.1146/annurev.aa.32.090194.001203}, \href
  {https://ui.adsabs.harvard.edu/abs/1994ARA&A..32..191W} {32, 191}

\bibitem[\protect\citeauthoryear{{Young}, {Alexander}, {Rosotti}  \&
  {Pinte}}{{Young} et~al.}{2022}]{Young2022}
{Young} A.~K.,  {Alexander} R.,  {Rosotti} G.,   {Pinte} C.,  2022, \mn@doi
  [\mnras] {10.1093/mnras/stac840}, \href
  {https://ui.adsabs.harvard.edu/abs/2022MNRAS.513..487Y} {513, 487}

\bibitem[\protect\citeauthoryear{{Yu}, {Teague}, {Bae}  \& {{\"O}berg}}{{Yu}
  et~al.}{2021}]{Yu2021}
{Yu} H.,  {Teague} R.,  {Bae} J.,   {{\"O}berg} K.,  2021, \mn@doi [\apjl]
  {10.3847/2041-8213/ac283e}, \href
  {https://ui.adsabs.harvard.edu/abs/2021ApJ...920L..33Y} {920, L33}

\bibitem[\protect\citeauthoryear{{Zhang}, {Blake}  \& {Bergin}}{{Zhang}
  et~al.}{2015}]{Zhang2015}
{Zhang} K.,  {Blake} G.~A.,   {Bergin} E.~A.,  2015, \mn@doi [\apjl]
  {10.1088/2041-8205/806/1/L7}, \href
  {https://ui.adsabs.harvard.edu/abs/2015ApJ...806L...7Z} {806, L7}

\makeatother
\end{thebibliography}

%%%%%%%%%%%%%%%%%%%%%%%%%%%%%%%%%%%%%%%%%%%%%%%%%%

%%%%%%%%%%%%%%%%% APPENDICES %%%%%%%%%%%%%%%%%%%%%

\appendix

\section{Formulae to compute molecular gas column density}
\label{sec:appendix1}

\subsection{Optical depth}
Prior to computing gas column density, the optical depths of the emission lines intended to be used needs to be determined, to ensure an optically thick tracer will be used to trace gas kinematic temperature whilst the tracer intended for computing column density is indeed optically thin. Under the assumption of local thermal equilibrium, optical depths ($\tau$) can be roughly estimated using the line ratio $R$ between two tracers close in frequency \citep[e.g.][]{Lyo2011, Kastner2018},
\begin{equation}
    \label{eq:line_ratio}
    R = \frac{T_{B}(\nu_{1})}{T_{B}(\nu_{2})} = \frac{1 - e^{-\tau_{\nu_{1}}}}{1 - e^{-\tau_{\nu_{2}}}} = \frac{1 - e^{-\tau_{\nu_{1}}}}{1 - e^{-\tau_{\nu_{1}}/X}},
\end{equation}
where $\nu_{i=1,2}$ are the rest frequencies of the lines. X is the molecular abundance ratio between the two lines.

\subsection{Molecular column density}
\label{app:CD}
The molecular column density (in units of  molecules per m$^{2}$) for an optically thin uniform slab is given by,
\begin{equation}
    \label{eq:col1}
    N_{\rm mol} = N_{u} \frac{Z}{2J + 1}\exp{\left[ \frac{E_u}{kT_{\rm ex}}\right]},
\end{equation}
where $J$=2, $k$ is the Boltzmann constant, $E_u$ is the energy of the upper state and $Z$ is the partition function of a linear molecule.

The population in the upper energy state, $N_{u}$, is given by \citep{Mangum2015},
\begin{equation}
    \label{eq:col2}
     N_{u} = f \frac{4\pi}{h\nu A_{ul}}  \frac{B_\nu(T_\mathrm{ex})}{B_\nu(T_\mathrm{ex}) - B_\nu(T_\mathrm{bg})} \int (I_\nu -  I_{\nu, \mathrm{bg}})\, \mathrm{d}v,
\end{equation}
where $A_{ul}$ is the Einstein coefficient for spontaneous emission, $h$ is Planck's constant, $\nu$ is the rest frequency, $T_\mathrm{ex}$ is the excitation temperature, $I_\nu - I_{\nu, \mathrm{bg}}$ is the specific intensity minus the background, \emph{e.g.} the quantity directly measured by the interferometer due to spatial filtering, and $f$ is the beam filling factor, where $f$=1 corresponds to the scenario where emission completely fills the beam.

\section{Channel maps}

\begin{figure*}
    \centering
    \includegraphics[width=\textwidth]{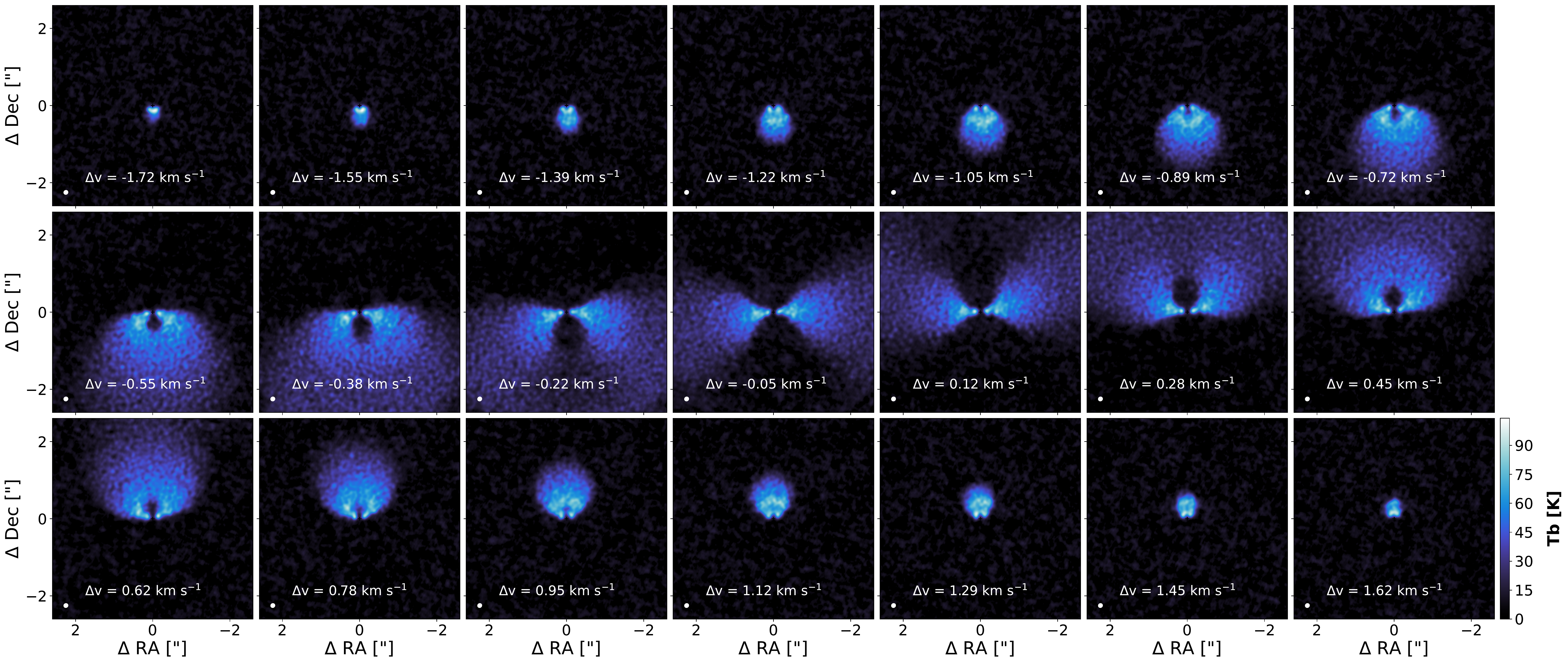}
    \caption{ALMA data: $^{12}$CO $J$=2-1 channel maps.}
    \label{}
\end{figure*}

\begin{figure*}
    \centering
    \includegraphics[width=\textwidth]{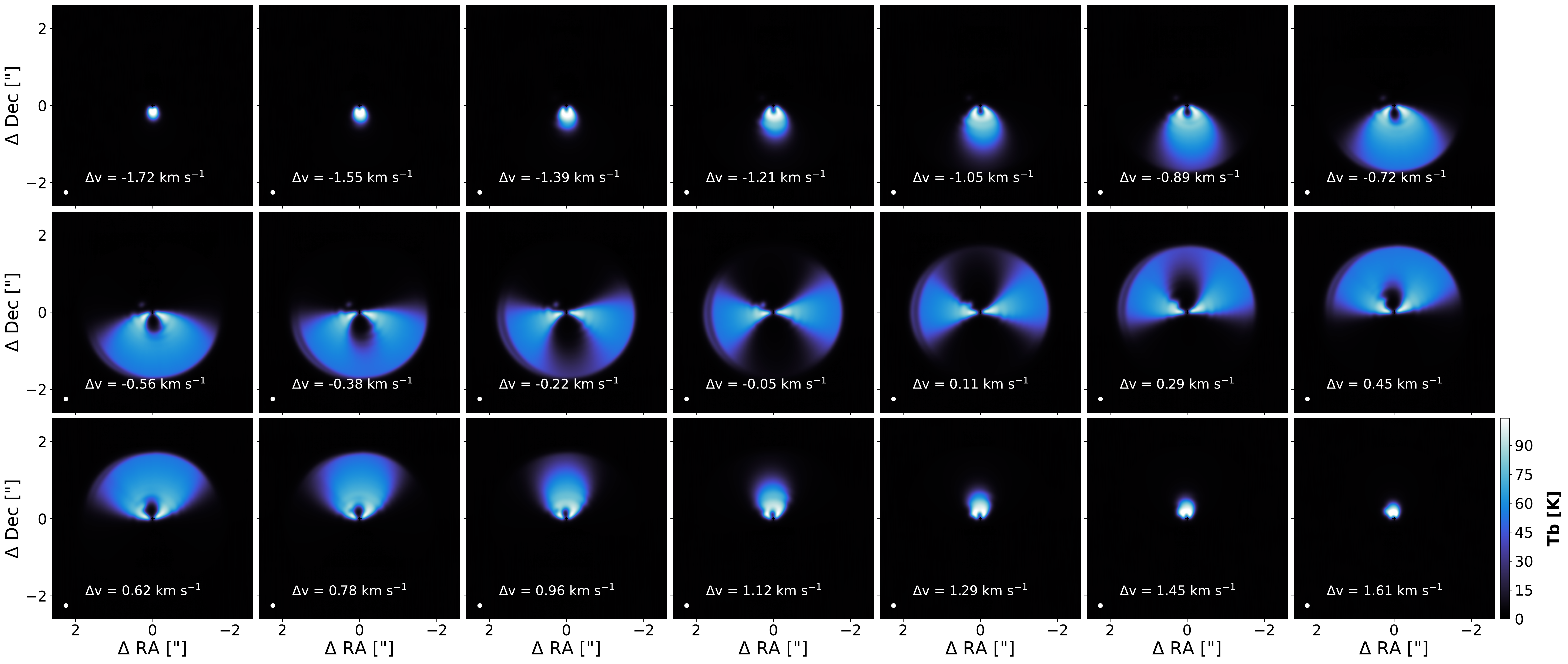}
    \caption{{\sc mcfost} synthetic data: $^{12}$CO $J$=2-1 channel maps with an embedded 10\mjup\,planet at 38\,au from disc centre.}
    \label{fig:B2}
\end{figure*}

\begin{figure*}
    \centering
    \includegraphics[width=\textwidth]{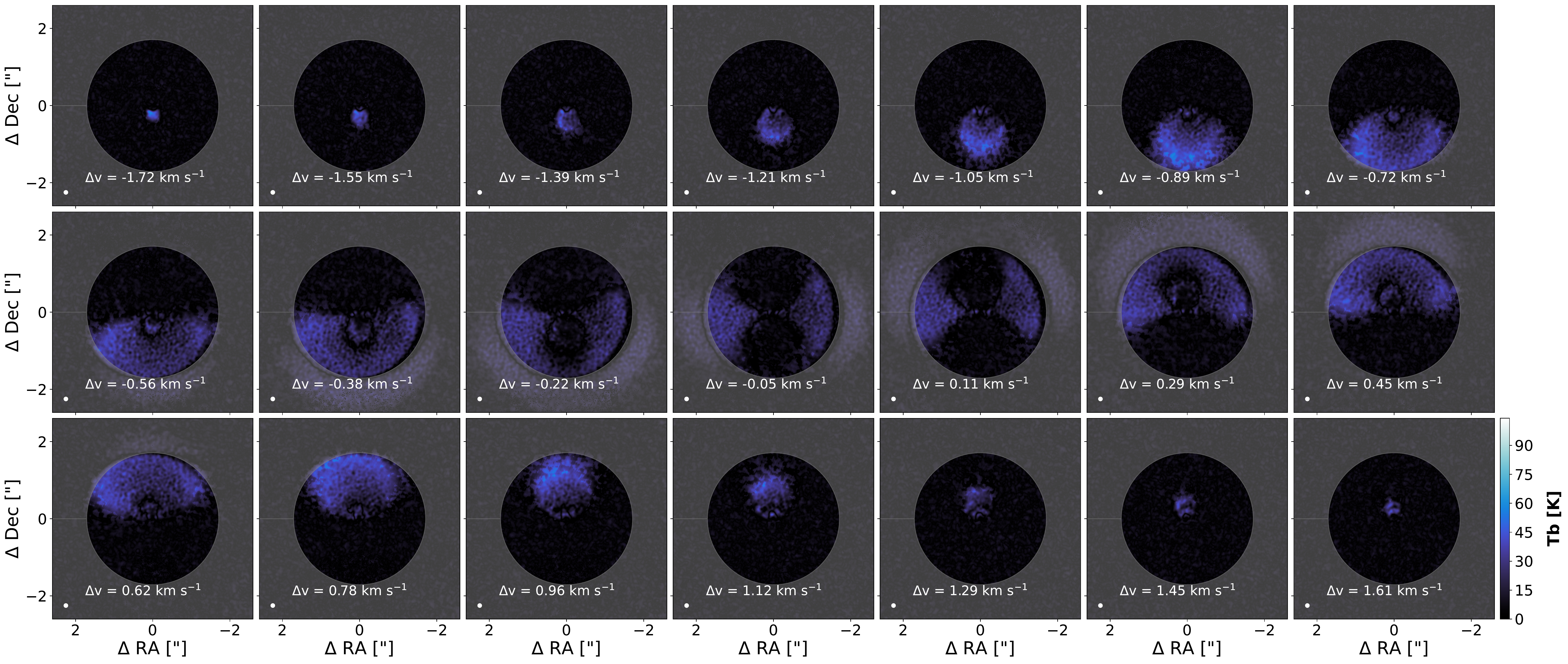}
    \caption{Residual $^{12}$CO $J$=2-1 channel maps: |ALMA data - {\sc mcfost} synthetic data|. The grey shaded area represents a mask at r > 200\,au; the outer radius of our synthetic data in Figure \ref{fig:B2}. Here we stress that non-zero brightness residuals should not be inferred as a proxy for the goodness of fit of the velocity fields.}
    \label{}
\end{figure*}

\begin{figure*}
    \centering
    \includegraphics[width=\textwidth]{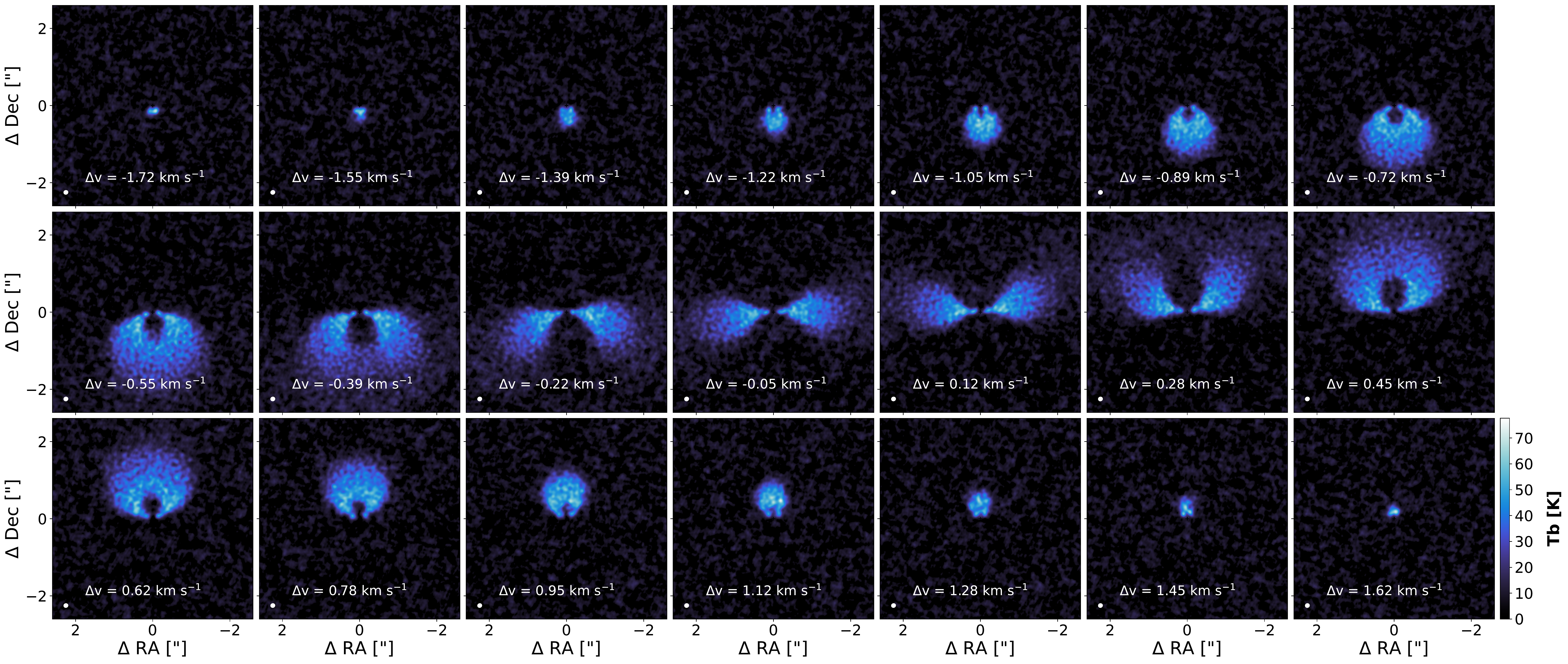}
    \caption{$^{13}$CO $J$=2-1 channel maps.}
    \label{}
\end{figure*}

\begin{figure*}
    \centering
    \includegraphics[width=\textwidth]{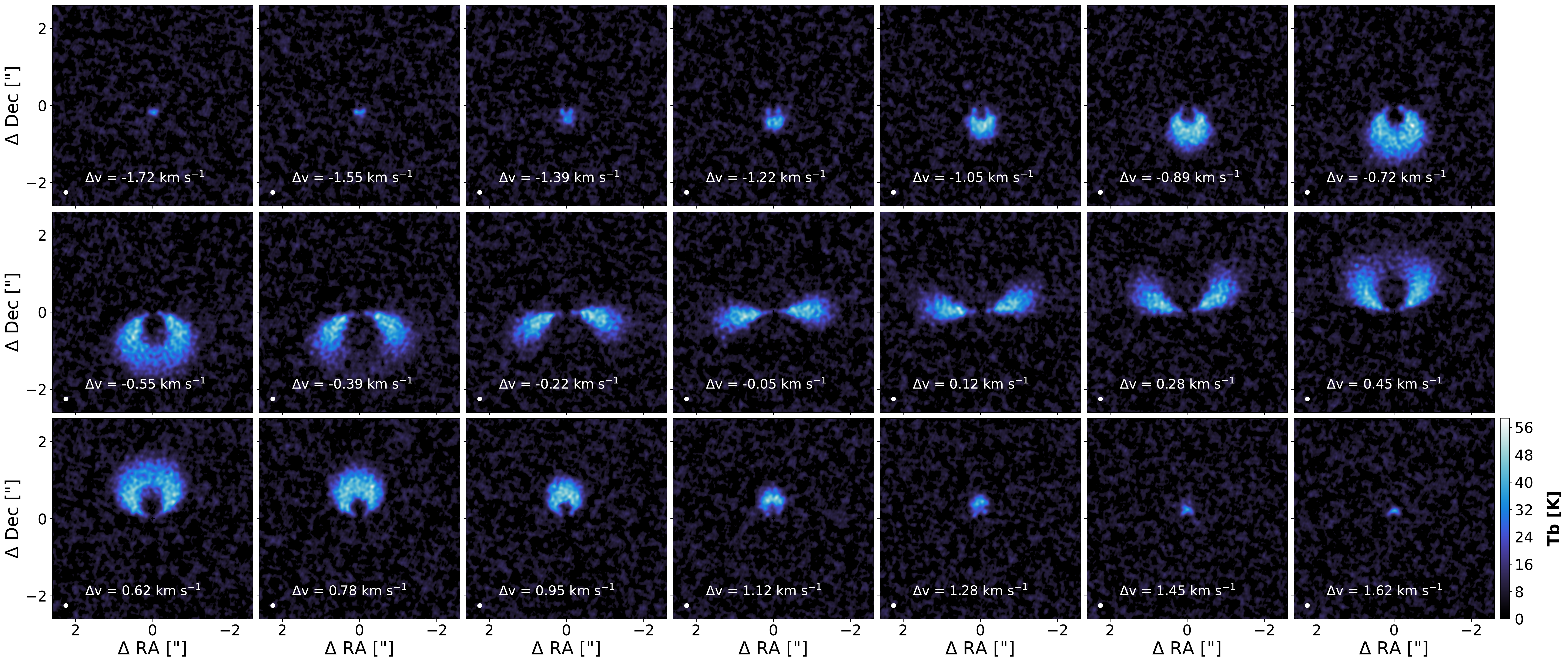}
    \caption{C$^{18}$O $J$=2-1 channel maps.}
    \label{}
\end{figure*}

\section{Incorporating uncertainties to the best fit parameters determined from {\sc eddy}}
\label{sec:eddy_uncertainty}

\begin{figure*}
    \centering
    \includegraphics[width=\textwidth]{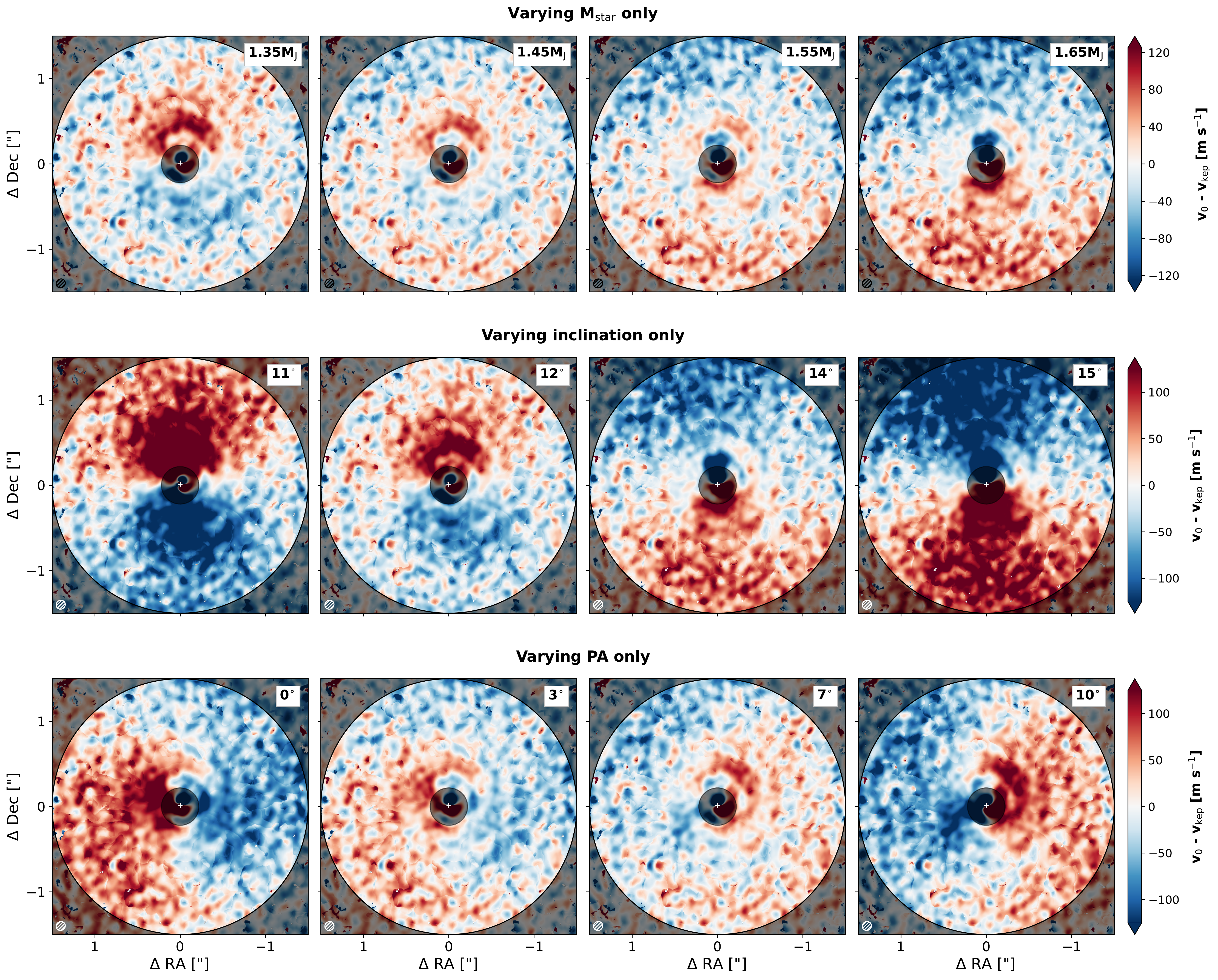}
    \caption{\textbf{Uncertainty series:} $^{12}$CO residual velocity maps for varying quantities of \msun, inclination and PA around the best fit values determined from the MCMC fitting in {\sc eddy}.}
    \label{fig:eddy_uncertainty}
\end{figure*}

%%%%%%%%%%%%%%%%%%%%%%%%%%%%%%%%%%%%%%%%%%%%%%%%%%

% Don't change these lines
\bsp	% typesetting comment
\label{lastpage}
\end{document}